\documentclass[aps,prb,twocolumn,superscriptaddress,longbibliography,floatfix]{revtex4-2}
\pdfoutput=1
\usepackage{braket}
\usepackage{graphicx}
\usepackage{dcolumn}
\usepackage{amsmath}
\usepackage{bm}
\usepackage{color}
\usepackage{hyperref}
\usepackage{multirow}
\usepackage{hhline}
\usepackage{xcolor}
\usepackage{makecell}
\usepackage{esint}

\usepackage[caption=false]{subfig}

\begin{document}

\title{{Nonlinear thermoelectric effects driven by \textcolor{black}{electron} quantum geometry}}
\author{Xu Yang}
\affiliation{Department of Physics, Ohio State University, Columbus, Ohio 43202, USA}
\affiliation{School of Physical and Mathematical Sciences, Nanyang Technological University, 637371, 
Singapore}
\date{\today}

\author{Brian Skinner}
\affiliation{Department of Physics, Ohio State University, Columbus, Ohio 43202, USA}
\date{\today}

\begin{abstract}
{Nonlinear thermoelectric effects offer the potential to enable new energy technologies, such as voltage-controlled thermal switching and thermoelectric rectification. 
In this work, we examine how quantum geometry \textcolor{black}{of the electron bands} gives rise to nonlinear thermoelectric responses in materials with suitable symmetries.} We derive a series of nonlinear thermoelectric effects governed by the Berry curvature dipole and the quantum metric dipole, respectively. Among them, we identify a particularly interesting quantized thermoelectric response that directly measures the total chirality of Weyl points below the Fermi level. For general nonlinear responses, we derive the nonlinear analogs of the Wiedemann-Franz law and Mott's formula. These provide a means to estimate the magnitude of nonlinear thermoelectric responses based on existing nonlinear Hall measurements. Our estimates suggest that these effects should be observable in several candidate materials, with In-doped Pb$_{1-x}$Sn$_x$Te standing out as the most promising. Our work offers new insights into the experimental study of quantum geometry through nonlinear thermoelectric measurements.
\end{abstract}

\maketitle

\section{Introduction}

{
Electrons moving through a solid material carry both electric current and heat. This dual role gives rise to thermoelectric effects in a conducting solid material, which allow for the conversion between electrical and thermal energy. The strength of thermoelectric effects is usually quantified by various linear response coefficients, such as the Seebeck and Nernst coefficients. Research into thermoelectric energy technologies generally aims to maximize the values of these coefficients (while maintaining relatively high electrical conductivity and low thermal conductivity) \cite{shakouri_recent_2011, yan_high-performance_2022, freer_realising_2020, zhang_thermoelectric_2015, fu_topological_2020, adachi_fundamentals_2025, skinner_design_2025}.}

{
Very recent work, on the other hand, has suggested the potential importance of \emph{nonlinear} thermoelectric effects \cite{zeng2020fundamental, varshney2024intrinsic, chen2016nonlinear, zeng2019nonlinear, zeng2021nonlinear, nakai2019thermalelectric, arisawa_observation_2024, nomoto_observation_2025}, in which a current response (electrical or thermal) is proportional to the product of an electric field and temperature gradient. Such nonlinear responses can be generated by a variety of effects that are extrinsic to the electron system, such as skew-scattering or side-jump scattering by impurities \cite{nagaosa2010anomalous}, or by chirality of the phonon system \cite{nomoto_observation_2025}. Here we focus instead on nonlinear thermoelectric effects that are intrinsic to the electron system, driven by the quantum geometry of the electronic band structure. Such effects have the potential to enable qualitatively new kinds of thermoelectric technologies, such as voltage-controlled heat switches \cite{wehmeyer_thermal_2017}, in which a transverse electric field controls a material's thermal conductivity, and thermoelectric rectifiers \cite{chang_solid-state_2006, li_thermal_2004}, in which an AC voltage generates a DC heat flux, or vice-versa. Further, the effects we consider here are controlled by the time scales associated with the electron system (e.g., the momentum relaxation time), so that they constitute time-reversal-symmetry breaking effects that can be controlled or switched at THz frequencies (as compared, say, to the much slower time scales associated with magnetic switching \cite{kirilyuk_ultrafast_2010}). At a more fundamental level, since the quantum geometric response is an intrinsic property of the electronic band structure, measurements of its magnitude provide a direct probe of the electronic quantum geometry.}

A fundamental quantity that captures the geometric structure of electronic wave functions is the quantum geometric tensor (QGT), defined in momentum space as:
\begin{align}
&T_{\alpha\beta}(\textbf{k})\equiv \braket{\partial_{k_{\alpha}}u_{\textbf{k}}|\partial_{k_{\beta}}u_{\textbf{k}}}-\braket{\partial_{k_{\alpha}}u_{\textbf{k}}|u_{\textbf{k}}}\braket{u_{\textbf{k}}|\partial_{k_{\beta}}u_{\textbf{k}}},
\end{align}
where $\ket{u_{\textbf{k}}}$ is the periodic part of the Bloch wave-function with wave vector $\textbf{k}$.

The QGT encodes both the quantum metric $g$ (real part) and the Berry curvature $\Omega$ (imaginary part) \cite{berry1984quantal,provost1980riemannian,xiao2010berry,rossi2021quantum,torma2023essay,verma2025quantum}. While the Berry curvature governs various topological and transport phenomena—including the quantum Hall effect\cite{thouless1982quantized,niu1985quantized}, anomalous Hall effect\cite{sundaram1999wave,haldane2004berry,nagaosa2010anomalous}, and chiral anomaly in Weyl semimetals\cite{zyuzin2012topological,son2013chiral}—the quantum metric influences physical properties such as superfluid weight in a flat band\cite{peotta2015superfluidity}, exciton spectra\cite{srivastava2015signatures}, and orbital magnetic susceptibilities\cite{gao2015geometrical,piechon2016geometric}. The ability to experimentally probe these geometric properties is therefore of great significance.

Among the most effective probes of quantum geometry are nonlinear Hall effects (NLHE), where an electric current emerges at second order in an applied electric field \cite{moore2010confinement,sodemann2015quantum,ma2019observation,kang2019nonlinear,du2021nonlinear,wang2021INH,liu2021INH,gao2023quantum,du2021nonlinear}. These effects arise in systems lacking inversion symmetry and can be driven by two primary mechanisms: (i) the Berry-curvature-dipole (BCD) induced NLHE, which stems from a momentum-space dipole in the Berry curvature, and (ii) the quantum-metric-dipole (QMD) induced NLHE, which originates from a similar dipole structure in the quantum metric. Crucially, these two mechanisms transform differently under time-reversal symmetry: BCD-induced NLHE has been observed in time-reversal symmetric systems, while QMD-induced NLHE emerges in $\mathcal{P}\mathcal{T}$-symmetric antiferromagnets. This difference in symmetry allows the two effects to be individually observed in experiments.

Despite the substantial theoretical and experimental progress made in studying the electronic NLHE, the exploration of nonlinear thermoelectric effects remains relatively underdeveloped. In this paper, we aim to fill this gap by formulating a unified theoretical framework for nonlinear thermoelectric effects induced by BCD and QMD, which offer an alternative pathway for probing quantum geometry in materials. Specifically, of all the different kinds of nonlinear thermoelectric effects\cite{chen2016nonlinear,zeng2019nonlinear,yu2019topological,nakai2019thermalelectric,zeng2020fundamental,zeng2021nonlinear,wang2022quantum,yamaguchi2024microscopic,varshney2024intrinsic}, we focus on the charge and heat currents resulting from the interplay between electric fields and temperature gradients: $j_e,j_Q\propto E( \nabla T)$. Using Boltzmann equations, we derive the expressions for the nonlinear thermoelectric effects induced by the BCD and the QMD. \textcolor{black}{While the expressions for the BCD-induced nonlinear thermoelectric coefficients were first derived in Ref.~\cite{nakai2019thermalelectric}, a key result of our work is the identification of a novel quantized nonlinear thermoelectric effect that directly measures the total Weyl chirality below the Fermi energy.} Remarkably, this effect is robust against band curvature and interaction effects, making it an advantageous tool for probing quantum geometry. We highlight SrSi$_2$ as a promising material for experimental observation of this effect. \textcolor{black}{In addition, we uncover a new nonlinear thermoelectric effect arising from the quantum metric dipole (QMD) and derive it microscopically. Building on these results, we present simple model calculations to support experimental investigations, demonstrating the behavior of BCD-induced and QMD-induced nonlinear thermoelectric responses with respect to doping and temperature.}

Finally, we employ Sommerfeld expansions to relate nonlinear thermoelectric effects to nonlinear Hall effects, enabling realistic estimates of thermoelectric responses based on existing NLHE data. This allows us to identify a wide range of promising candidate materials, including In-doped Pb$_{1-x}$Sn$_x$Te, TaAs, and MnBi$_2$Te$_4$, with In-doped Pb$_{1-x}$Sn$_x$Te standing out due to its sizable change in thermal conductivity under an electric field, readily accessible with current experimental techniques. A comprehensive table of realistic estimates for the nonlinear thermoelectric coefficients is provided for these materials, establishing a concrete benchmark for future experimental investigations.

\section{Theoretical formulation}\label{sec:theory}
To set the stage for detecting quantum geometry via second-order nonlinear thermoelectric effects, we begin with the nonlinear electrical response and its connection to quantum geometry. The second-order conductivity tensor $\sigma_{abc}$ relates the induced current to the applied electric fields via $j_e^a = \sigma_{abc} E_b E_c$, with repeated indices summed over. A broad class of second-order responses arises from disorder-mediated mechanisms, such as side-jumps and skew scattering, and is collectively termed extrinsic effects. Their connection to quantum geometry, as well as the discussion of their thermoelectric analogs, depends sensitively on the underlying scattering processes and is thus left for future work. The intrinsic mechanisms, on the other hand, are determined solely by the properties of the quantum state. They include both Ohmic and Hall contributions, analogous to the linear response case. The Ohmic part, second order in relaxation time $\tau$, accounts for dissipative effects and arises from symmetrizing over all permutations of $a,b,c$. As a natural extension of the linear Drude term, it is referred to as the nonlinear Drude effect. It typically involves energy derivatives and is generally unrelated to quantum geometry. The Hall part is then defined as the total response minus the Ohmic part and is closely linked to quantum geometry\cite{tsirkin2022separation}. More specifically, nonlinear Hall effects enable direct probing of the Berry curvature dipole and the quantum metric dipole. Based on their $\tau$-dependence, nonlinear Hall effects can be classified into the following two types:
\begin{enumerate}
\item The Berry curvature dipole (BCD)-induced nonlinear Hall effect, with coefficient denoted as $\sigma_{BCD}$, is linear in the relaxation time $\tau$. This effect arises only when $\mathcal{P}\mathcal{T}$ symmetry is broken.
\item The intrinsic nonlinear Hall effect, with coefficient denoted as $\sigma_\text{INH}$, is independent of the relaxation time $\tau$ and occurs only when $\mathcal{T}$ symmetry is broken. We refer to this as the quantum metric dipole (QMD)-induced nonlinear Hall effect, a terminology that will be justified in Sec.~\ref{sec:qmd_introduction}.
\end{enumerate}
While these two mechanisms for nonlinear Hall effects generally coexist in non-centrosymmetric systems, their distinct transformation properties under time-reversal symmetry allow us to isolate them when the system possesses either $\mathcal{T}$ or $\mathcal{P}\mathcal{T}$ symmetry.

\begin{figure}[t]
    \centering
    \includegraphics[width=0.50\textwidth]{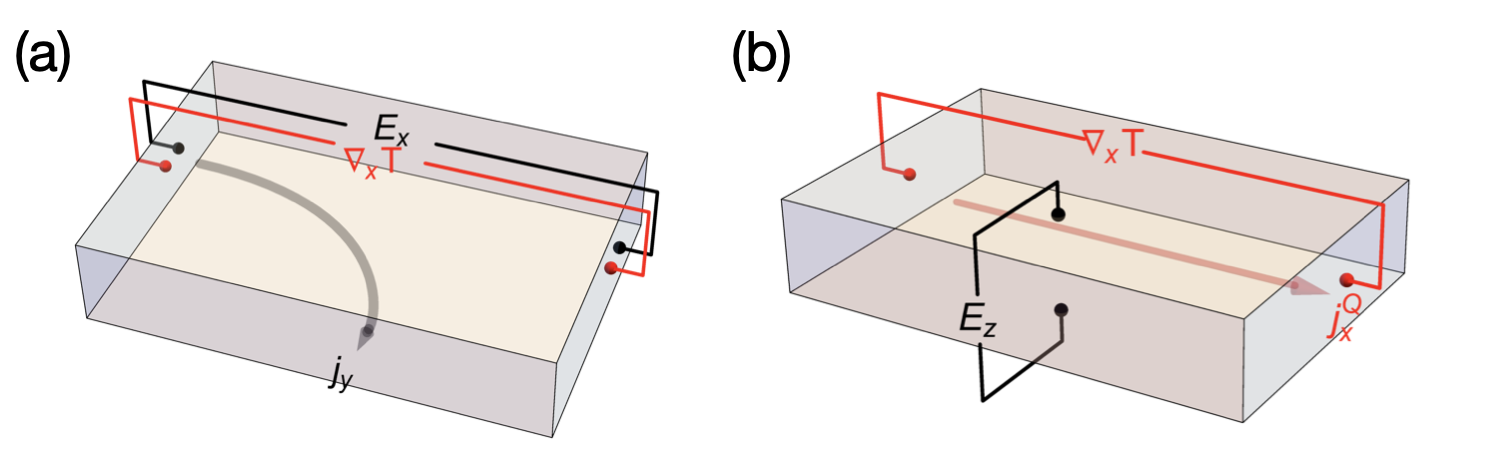}
    \caption{Schematic illustrations of two instances of the nonlinear thermoelectric effect. (a) Measuring the change in $\sigma_{xy}$ induced by a temperature gradient $\nabla_x T$, from which we can extract $\hat{L}^{112}_{yxx}+\hat{M}^{112}_{yxx}$.  (b) Measuring the change in $\kappa_{xx}$ induced by an electric field $E_z$, from which we can determine $\hat{L}^{212}_{xzx}+\hat{M}^{212}_{xzx}$.}
    \label{fig:measurement}
\end{figure}

These considerations motivate us to seek analogs of nonlinear Hall effects that probe quantum geometry through nonlinear thermoelectric responses. In particular, we focus on nonlinear thermoelectric effects driven by the Berry curvature dipole 
and the quantum metric dipole. 
\textcolor{black}{The most general second-order nonlinear responses contain components such as $j_Q\propto EE$ and $j_e \propto (\nabla T)(\nabla T)$. In this work, however, we restrict our attention to mixed responses $j_e,j_Q\propto E(\nabla T)$. This choice is motivated by two reasons. First, the mixed nonlinear thermoelectric effect is experimentally accessible either as an electric-field-induced correction to the thermal conductivity or as a temperature-gradient-induced correction to the electric conductivity (see Sec.~\ref{sec:experimental_meaning}). Second, as we shall see in Sec.~\ref{sec:quantized_response}, the nonlinear thermoelectric effect gives rise to a quantized response arising from the interplay between electric and thermal probes, which is absent in purely electric or purely thermal channels. More specifically, the two types of nonlinear thermoelectric effects under consideration are}:
\begin{enumerate}
    \item 
Charge current induced by the interplay between the electric field and the temperature gradient, $j_e\propto E ( \nabla T)$. Following the expression in Ref.\onlinecite{nakai2019thermalelectric}, we write 
\begin{align}
&j_e^a=\sum\limits_{b,c}(\hat{L}^{112}_{abc}+\hat{M}^{112}_{abc})E_b\nabla_cT,
\end{align}
where $L$ and $M$ are tensors that describe the BCD and QMD induced nonlinear thermoelectric effects, respectively, and the first index 1 of the superscript indicates the electric current, the second index 1 indicates the electric field, and the third index 2 indicates the temperature gradient, a convention we use below. Because a temperature gradient $\nabla T$ transforms under symmetry in exactly the same way as an electric field, the symmetry discussion for the nonlinear Hall effects continues to hold. Therefore both $L$ and $M$ vanish for a centrosymmetric system,  $\hat{L}^{112}$ vanishes if the system has $\mathcal{P}\mathcal{T}$ symmetry, and $\hat{M}^{112}$ vanishes for $\mathcal{T}$-symmetric system. For a generic system, the nonlinear thermoelectric effects have contributions from both $\hat{L}^{112}$ and $\hat{M}^{112}$, but these contributions can be singled out if either $\mathcal{P}\mathcal{T}$ or $\mathcal{T}$ symmetries are present. In Fig.\ref{fig:measurement}(a) we illustrate an instance of these nonlinear thermoelectric effects.

\item 
Heat current induced by the interplay between the electric field and the temperature gradient, ${j_Q\propto E ( \nabla T)}$. We can express these effects as:
\begin{align}
&j_Q^a=\sum\limits_{b,c}(\hat{L}_{abc}^{212}+\hat{M}^{212}_{abc})E_b\nabla_cT,
\end{align}
where the first index 2 of the superscript indicates the heat current, and all other conventions follow from the previous discussion. The heat current $j_Q$ transforms under symmetry in exactly the same way as the electric current and therefore the symmetry properties of $\hat{L}^{212}, \hat{M}^{212}$ are exactly the same as $\hat{L}^{112}, \hat{M}^{112}$ and $\sigma_{{BCD}}, \sigma_{{QMD}}$ listed above. In Fig.\ref{fig:measurement}(b) we illustrate an instance of these nonlinear thermoelectric effects.
\end{enumerate}

Below, we derive the expressions for nonlinear thermoelectric responses induced by the BCD and the QMD based on the semiclassical equations of motion. The contribution arising from the BCD was first presented in Ref.~\cite{nakai2019thermalelectric}, and we briefly recapitulate the result and proceed to discuss its physical consequence. In contrast, the nonlinear thermoelectric effect induced by QMD constitutes a novel contribution, for which a detailed derivation is in Sec.\ref{sec:qmd_introduction}.

\subsection{Berry-curvature-dipole induced nonlinear thermoelectric effects}
In this section, we first provide a brief summary of the semiclassical dynamics method that will be used to derive the expressions for nonlinear thermoelectric effects induced by both the Berry curvature dipole (BCD) and the quantum metric dipole (QMD). We then present the expressions for the BCD-induced nonlinear thermoelectric coefficients, namely $\hat{L}^{112}$ and $\hat{L}^{212}$. The derivation was first given in Ref.~\cite{nakai2019thermalelectric}. Here we simply reproduce the result and refer the reader to the Appendix~\ref{appendix:derivation} for technical details. \textcolor{black}{We note that the same thermoelectric responses can, in principle, also be derived within a Kubo formalism using the thermal length-gauge approach for nonlinear thermal transport~\cite{wang2022quantum}, which admits a natural diagrammatic interpretation analogous to nonlinear optical responses~\cite{parker2019diagrammatic}.}

Within the framework of semiclassical dynamics, one can construct wave packets of Bloch electrons that possess both a well-defined center-of-mass position and momentum.
In the presence of an $E$ field, the position $\textbf{r}$ and the wave vector $\hbar \textbf{k}$ of an electron wave-packet satisfies:
\begin{equation}
    \begin{split}
        &\dot{\textbf{r}}=\frac{1}{\hbar}\nabla_{\textbf{k}}\varepsilon-\dot{\textbf{k}}\times\bm{\Omega},\\
        &\hbar\dot{\textbf{k}}=-e\bm{E},\\
    \end{split}
\end{equation}
where the second term on the RHS in the first equation, $-\dot{\textbf{k}}\times \bm{\Omega}$, is the anomalous velocity term \cite{karplus1954hall,sundaram1999wave,xiao2010berry} that arises from the semiclassical wave-packet dynamics, and $\bm{\Omega}\equiv i\bra{\nabla_{\textbf{k}}u_{\textbf{k}}}\times \ket{\nabla_{\textbf{k}}u_{\textbf{k}}}$ is the Berry curvature that acts like a $k$-space ``Lorentz force." The Berry curvature $\bm{\Omega}$ is the curl of the Berry connection $\bm{\mathcal{A}}$, so that $\bm{\Omega}\equiv\nabla_{\textbf{k}}\times \bm{\mathcal{A}}$ with $\bm{\mathcal{A}}\equiv -i\bra{u_{\textbf{k}}}\nabla_{\textbf{k}}\ket{u_{\textbf{k}}}$. It is worth commenting that all the quantum geometric effects studied in this work originate from this anomalous velocity term.

From the Boltzmann equation, the non-equilibrium distribution function $g_{\textbf{r},\textbf{k}}$ satisfies the following equation:
\begin{equation}
    \frac{\partial g_{\textbf{r},\textbf{k}}}{\partial t}+\dot{\textbf{r}}\cdot \nabla_{\textbf{r}}g_{\textbf{r},\textbf{k}}+\dot{\textbf{k}}\cdot\nabla_{\textbf{k}}g_{\textbf{r},\textbf{k}}=-\frac{g_{\textbf{r},\textbf{k}}-f}{\tau},
\end{equation}
where ${f = 1/[\exp((\varepsilon_{\textbf{k}}-\mu)/k_BT) + 1]}$ is the equilibrium Fermi-Dirac distribution function with chemical potential $\mu$ and we have made the relaxation time approximation with $\tau$ the relaxation time. If we assume the electric field is applied at frequency $\omega$, so that $\bm{E}(t)=Re[e^{i\omega t}\bm{E}_0]$, we can solve the Boltzmann equation to give the following expression for the electric current\cite{nakai2019thermalelectric} (see Appendix.\ref{appendix:derivation}): 
\begin{equation}
    \begin{split}
\bm{j}_e = & \int[dk](-e\bm{v})\frac{1}{i\omega+1/\tau}\frac{e}{\hbar}[(\bm{E}\times \bm{\Omega})\cdot \nabla T]\frac{(\varepsilon-\mu)}{T}\partial_{\varepsilon}f + \\
& (-e)\int[dk]\tau[\bm{v}\cdot \nabla T]\frac{(\varepsilon-\mu)}{T}(\partial_{\varepsilon}f)(\frac{e}{\hbar}\bm{E}\times\bm{\Omega}),  
    \end{split}
\end{equation}
where $[dk]$ denotes a shorthand for the $D$-dimensional momentum integration measure $\frac{d^Dk}{(2\pi)^D}$. From this expression we can obtain the coefficient $\hat{L}_{abc}^{112}$\footnote{We note that the first terms of the RHS of Eqs.(42) and (55) in Ref.\cite{nakai2019thermalelectric} do not appear to incorporate the minus sign associated with the temperature gradient as defined in Eq.(40). After correcting these, their expressions are in exact agreement with our expressions Eq.\eqref{eq:L112} and Eq.\eqref{eq:L212}.}:
\begin{align}\label{eq:L112}
& \hat{L}_{abc}^{112}=\frac{e^2\tau}{\hbar}\int [dk]\frac{(\varepsilon-\mu)}{T}\left[ \frac{\epsilon_{bcd}}{i\omega\tau+1}(\partial_af)\Omega_d-\epsilon_{abd}(\partial_cf)\Omega_d \right].
\end{align}
Similarly, we can compute the heat current induced by a temperature gradient and an electric field. The result is:
\begin{align}\label{eq:L212}
& \hat{L}_{abc}^{212}=-\frac{e\tau}{\hbar}\int [dk]\frac{(\varepsilon-\mu)^2}{T}\left[\frac{\epsilon_{bcd}}{i\omega\tau+1}(\partial_af)\Omega_d-\epsilon_{abd}(\partial_cf)\Omega_d \right].
\end{align}

\begin{widetext}
The Berry curvature dipole (BCD) is defined as $D_{\text{Berry},bd}(\varepsilon_F)=\hbar\int [dk]v_b\Omega_d\delta(\varepsilon-\varepsilon_F)$. Then we can re-express the above formulas in terms of weighted averages over Berry curvature dipoles:
\begin{align}\label{eq:bcdintegration1}
    &\hat{L}^{112}_{abc}=\frac{e^2\tau}{\hbar^2}\int d\varepsilon\frac{(\varepsilon-\mu)}{T}(\partial_{\varepsilon}f) \left[\frac{\epsilon_{bcd}}{i\omega\tau+1}D_{\text{Berry},ad}(\varepsilon)-\epsilon_{abd}D_{\text{Berry},cd}(\varepsilon) \right],\\
   \label{eq:bcdintegration2} &\hat{L}^{212}_{abc}=-\frac{e\tau}{\hbar^2}\int d\varepsilon\frac{(\varepsilon-\mu)^2}{T}(\partial_{\varepsilon}f)\left[\frac{\epsilon_{bcd}}{i\omega\tau+1}D_{\text{Berry},ad}(\varepsilon)-\epsilon_{abd}D_{\text{Berry},cd}(\varepsilon) \right].
\end{align}

\textcolor{black}{At low temperature, we can use the Sommerfeld expansion to relate $\hat{L}^{112}$ and $\hat{L}^{212}$ to the Berry curvature dipole (which is defined at zero temperature):
\begin{align}\label{eq:L_Mottformula}
&\hat{L}^{112}_{abc}=-\frac{e^2}{\hbar^2}\frac{\pi^2}{3}(k_B^2T)\left[\frac{\epsilon_{bcd}\tau}{i\omega\tau+1}\frac{dD_{\text{Berry},ad}(\mu)}{d\mu}|_{\mu=\varepsilon_F} -\epsilon_{abd}\tau\frac{dD_{\text{Berry},cd}(\mu)}{d\mu}|_{\mu=\varepsilon_F} \right]+\mathcal{O}(T^3),\\
&\label{eq:L_WiedemannFranzlaw}
   \hat{L}^{212}_{abc}=\frac{e}{\hbar^2}\frac{\pi^2}{3}(k_B^2T)\left[\frac{\epsilon_{bcd}\tau}{i\omega\tau+1}D_{\text{Berry},ad}(\varepsilon_F) -\epsilon_{abd}\tau D_{\text{Berry},cd}(\varepsilon_F)\right]+\mathcal{O}(T^3),
\end{align}
where in the derivation we have also replaced $\mu(T)$ by Fermi energy at zero temperature $\varepsilon_F$ using $\mu=\varepsilon_F-\frac{\pi^2}{6}\frac{g'(\varepsilon_F)}{g(\varepsilon_F)}(k_BT)^2$\cite{Ashcroft76}, with $g(\varepsilon)$ being the density of states.}

These relations are reminiscent of Mott's formula and the Wiedemann–Franz law in linear response. As we show in Sec.\ref{sec:discussion} below, the Sommerfeld expansion provide a useful guide for experiments.

\textcolor{black}{Finally, let us comment on the validity of the constant relaxation time approximation employed throughout this work. In general, the relaxation time possesses momentum dependence and is determined by: $\tau_{\mathbf{k}}^{-1} = \sum_{\mathbf{k}'} W_{\textbf{k}\textbf{k}'}\mathcal{F}(v_{\textbf{k}},v_{\textbf{k}'})$, where $W_{\textbf{k},\textbf{k}'}$ is the scattering matrix element and $\mathcal{F}(v_{\textbf{k}},v_{\textbf{k}'})$ is a geometric factor depending on quasiparticle velocities. For instance, in an isotropic system, this factor simplifies to $\mathcal{F}=1-\text{cos}(\theta_{\textbf{k},\textbf{k}'})$, where $\theta_{\textbf{k},\textbf{k}'}$ is the angle between the initial and final wavevectors \cite{ziman2001electrons}. The energy dependence $\tau(\epsilon)$ is typically weak for a single, simply connected Fermi surface, where only states in a narrow window around the Fermi level contribute and $\tau_{\textbf{k}}$ is well approximated by its Fermi-surface value. A pronounced momentum-dependent $\tau_{\textbf{k}}$ arises only when (i) multiple symmetry-unrelated Fermi pockets introduce distinct, independent scattering channels—a scenario we discuss in Sec.~\ref{sec:quantized_response}—, or (ii) the underlying scattering mechanism itself is highly anisotropic, which we leave for a model-dependent study.}

\textcolor{black}{Notably, the scattering matrix elements $W_{\textbf{k},\textbf{k}'}$ can carry intrinsic quantum metric information via overlaps of Bloch states at different momenta. However, integrating over $\textbf{k}'$ compresses this geometric information into the scalar $\tau_{\textbf{k}}$ without generating new tensor structures. This smooth and positive-definite scalar modulation is easily distinguished from the intrinsic Berry curvature dipole contribution, which is a highly concentrated, symmetry-constrained rank-2 tensor that can alternate in sign.}
\end{widetext}
\subsection{Quantum-metric-dipole induced nonlinear thermoelectric effects}\label{sec:qmd_introduction}
We now consider quantum-metric-dipole (QMD) induced nonlinear thermoelectric effects, defined by the tensors $\hat{M}^{112}$ and $\hat{M}^{212}$, which are intrinsic effects independent of the relaxation time $\tau$. These effects are the thermoelectric analogs of the intrinsic nonlinear Hall effect\cite{xiao2006thermoelectric,gao2014qm,wang2021INH,liu2021INH}. The derivation of the expressions for $\hat{M}^{112},\hat{M}^{212}$ are entirely new, so we will explain it in detail here.


Let's derive the coefficient $\hat{M}^{112}$ from the anomalous Nernst effect, which is characterized by the generation of transverse electric current in response to a temperature gradient \cite{smrcka1977transport,qin2011energy,xiao2006thermoelectric,bergman2010theory}:
\begin{equation}\label{eq:linear_nernst}
    \bm{j}_e=-\frac{\nabla T}{T}\times\frac{e}{\hbar}\int[dk]\bm{\Omega}[(\varepsilon-\mu)f-k_BT\text{ln}(1-f)].
\end{equation}
This equation gives the Nernst conductivity $\alpha$, defined by $j_{e,i} = \alpha_{ij} \nabla_j T)$.

Equation \eqref{eq:linear_nernst} can be understood in the following way (we follow the quick derivation in Ref.~\cite{zhang2008anomalous}). First let's consider the coefficient $\bar{\alpha}$ defined as $j_{Q,i}=\bar{\alpha}_{ij}E_j$, which characterizes the transverse heat current in response to an electric field. The electric-field-induced transverse heat current can be represented as the entropy density multiplied by the anomalous velocity. (Since this component of the heat current is perpendicular to the electric field, it has no associated dissipation and is described by the thermodynamics of reversible processes, which dictates that the energy current is equal to $T$ times the entropy current \cite{Jay-Gerin_1974}.) This expression gives 
\begin{equation}
    \bm{j}_Q=\frac{e}{\beta\hbar}\int[dk](\bm{E}\times \bm{\Omega})s(\textbf{k}),
\end{equation} 
where $s(\textbf{k})=-f_{\textbf{k}}\text{log}(f_{\textbf{k}})-(1-f_{\textbf{k}})\text{log}(1-f_{\textbf{k}})$ is the entropy of electrons with wave vector $\textbf{k}$. We then use the Onsager relation $\alpha=\bar{\alpha}/T$ to obtain the Nernst conductivity, which gives us Eq.~\eqref{eq:linear_nernst}.

This derivation of Eq.~\eqref{eq:linear_nernst} assumes, implicitly, that the Berry curvature is unmodified by the electric field. But the presence of an electric field generically modifies the Berry connection $\bm{\mathcal{A}}$, and therefore also the Berry curvature $\bm{\Omega}$. To linear order in field, we can write 
\begin{align}\label{eq:BCP}
\mathcal{A}'_a =\mathcal{A}_a+eG_{ab}E_b,
\end{align}
where the physical Berry connection is denoted $\bm{\mathcal{A}}'$, and $\mathcal{A}_a\equiv -i\bra{u_0}\nabla_{k_a}\ket{u_0}$, with the subscript $0$ indicating the band index of the electronic wave-function under consideration. $G_{ab}$ is termed the Berry connection polarizability tensor in literature\cite{gao2014qm,liu2022berry}. The physical origin of $G_{ab}$ is the first-order correction to the eigenstate $\ket{u_0}$ of the Hamiltonian when perturbed by an electric field, and hence it can be conveniently calculated as:
\begin{align}\label{eq:BCP_expression}
G_{ab}=2\text{Re}\sum\limits_{n\neq 0}\frac{(\mathcal{A}_a)_{0n}(\mathcal{A}_b)_{n0}}{\varepsilon_0-\varepsilon_n},
\end{align}
where $(\mathcal{A}_a)_{0n}\equiv -i\bra{u_0}\nabla_{k_a}\ket{u_n}$ is the inter-band Berry connection between band $0$ and band $n$ (and $(\mathcal{A}_a)_{n0}$ is similarly defined). It is worth noting that although $\mathcal{A}_a$ and $\mathcal{A}'_a$ are gauge-dependent, the first order correction $eG_{ab}E_b$ is gauge invariant, as can also be seen from the explicit expression for $G$ given above (c.f. Ref.\cite{gao2014qm}). \textcolor{black}{Note that only the inter-band Berry connection appears in the Berry connection polarizability, with $\bm{\mathcal{A}}_{nm} = \frac{\bm{v}_{nm}}{i(\varepsilon_n-\varepsilon_m)}$ for $n \neq m$, where $\bm{v}_{nm}(\textbf{k})\equiv\frac{1}{\hbar}\bra{u_{n\textbf{k}}}\nabla_{\textbf{k}}H(\textbf{k})\ket{u_{m\textbf{k}}}$ is the velocity matrix element. Since velocity matrix elements and band energies are gauge invariant, this form makes the gauge invariance of the Berry connection polarizability, Eq.~\eqref{eq:BCP_expression}, more explicit.}

Inserting $\bm{\Omega}'=\nabla\times \bm{\mathcal{A}}'$ and Eq.\eqref{eq:BCP} into Eq.\eqref{eq:linear_nernst} and extracting the part that is proportional to $(\nabla T ) \vec{E}$, we arrive at: 
\begin{align}
    \bm{j}_{e}=-\frac{\nabla T}{T}\times e^2\int[dk] [(\textbf{G}\cdot \bm{E})\times \bm{v}](\varepsilon-\mu)\partial_\varepsilon f,
\end{align}
where the bold symbol $\textbf{G}$ indicates its tensorial nature. From it we get:
\begin{align}\label{eq:M112}
    \hat{M}_{abc}^{112}=-\frac{e^2}{ T}\int[dk](G_{ab}v_c-G_{cb}v_a)(\varepsilon-\mu)\partial_{\varepsilon}f.
\end{align}
The coefficient $\hat{M}^{212}$ can similarly be derived starting from the expression of the linear thermal Hall response and considering the first-order correction to the Berry curvature in an electric field. 

The thermal Hall response coefficient has been widely studied in the literature, and its derivation involves techniques beyond the scope of this work \cite{luttinger1964theory, cooper1997thermoelectric, qin2011energy,shitade2014heat, kapustin2020thermal}. Here, we simply state the result and refer interested readers to the relevant literature:
\begin{equation}
    \begin{split}
   &\bm{j}_Q=-\frac{k_B^2T}{\hbar}\nabla T\times \int [dk]\bm{\Omega} \Big[
 \frac{\pi^2}{3}+\beta^2(\varepsilon-\mu)^2f\\
&-\text{ln}^2(1-f)-2\text{Li}_2(1-f) \Big]
,     
    \end{split}
\end{equation}
where $\text{Li}_2(x)$ is the dilogarithm function.

In the presence of an electric field, the Berry curvature receives a correction that is first order in electric field. Once again utilizing Eq.\ref{eq:BCP} and replacing $\bm{\Omega}$ by $\bm{\Omega}'\equiv \nabla\times \bm{\mathcal{A}}'$, the desired nonlinear response originating from the interplay between $\nabla T$ and $\vec{E}$ can be written as:
\begin{align}
    \bm{j}_{Q}=-e\frac{\nabla T}{T}\times \int[dk] \left[ (\textbf{G}\cdot \bm{E})\times \bm{v} \right](\varepsilon-\mu)^2\partial_\varepsilon f,
\end{align}
which yields:
\begin{align}\label{eq:M212}
    \hat{M}_{abc}^{212}=-\frac{e}{ T}\int[dk](G_{ab}v_c-G_{cb}v_a)(\varepsilon-\mu)^2\partial_{\varepsilon}f.
\end{align}

Similar to the case of BCD-induced nonlinear thermoelectric effects, we can utilize the Sommerfeld expansion to relate the $\hat{M}^{112}$ and $\hat{M}^{212}$ to the intrinsic nonlinear Hall coefficient at zero temperature. To this end, we recall the definition of the intrinsic nonlinear Hall coefficient \cite{wang2021INH,liu2021INH}:
\begin{equation}\label{eq:sigma_INH}
    \sigma_{\text{INH},abc}(\mu)=-e^3\int [dk](G_{ab}v_c-G_{bc}v_a)\partial_{\varepsilon}f.
\end{equation}

We can then express the intrinsic nonlinear thermoelectric coefficients as a weighted integral of $\sigma_{\text{INH},abc}$ over different chemical potentials at zero temperature:
\begin{align}\label{eq:QMD_integrations1}
    &\hat{M}_{abc}^{112}(\mu,T)=-\frac{1}{eT}\int d\varepsilon (\varepsilon-\mu)(\partial_{\varepsilon}f)\sigma_{{\text{INH}},abc}(\varepsilon,T=0),\\ \label{eq:QMD_integrations2}
    &\hat{M}_{abc}^{212}(\mu,T)=-\frac{1}{e^2T}\int d\varepsilon (\varepsilon-\mu)^2(\partial_{\varepsilon}f)\sigma_{\text{INH},abc}(\varepsilon,T=0),
\end{align}

Therefore at low temperature we can once again use the Sommerfeld expansion to relate $\hat{M}^{112}$ and $\hat{M}^{212}$ to $\sigma_{\text{INH}}$:
\textcolor{black}{
\begin{align}\label{eq:M_Mott_formula}
    &\hat{M}_{abc}^{112}(\mu,T)=\frac{\pi^2}{3}\frac{k_B^2T}{e}\frac{d\sigma_{{\text{INH}},abc}(\mu,T=0)}{d\mu}|_{\mu=\varepsilon_F}+\mathcal{O}(T^3),
\end{align}
\begin{align}\label{eq:M_Wiedemann-Franz_Law}
    &\hat{M}_{abc}^{212}(\mu,T)=\frac{\pi^2}{3} \left(\frac{k_B}{e} \right)^2T\sigma_{{\text{INH}},abc}(\varepsilon_F,T=0)+\mathcal{O}(T^3).
\end{align}}

Let us now comment on the quantum geometrical meaning of the intrinsic nonlinear thermoelectric response. 
First, we recast the quantum metric $g_{\alpha, \beta} \equiv \text{Re}(T_{\alpha, \beta})$ for a band $0$ in terms of the inter-band Berry connection:
\begin{align}
& g_{ab}=\text{Re}\sum\limits_{n\neq 0}[(\mathcal{A}_a)_{0n}(\mathcal{A}_b)_{n0}].
\end{align}
Comparing with the definition of Berry connection polarizability Eq.\eqref{eq:BCP_expression}, we find that the only difference is the energy denominator. In the special case of a 2-band system labeled by band indices 0 and 1, we have a direct relation: $G_{ab}^{\text{2-band}}=2g_{ab}^{\text{2-band}}/(\varepsilon_0-\varepsilon_1).$

If the system consists of multiple bands, then in general there is no direct relation between $G_{ab}$ and $g_{ab}$. But if the quantity $\text{Re}[(\mathcal{A}_a)_{0n}(\mathcal{A}_b)_{n0}]$ only takes significant values for band $n=1$ that is close to band $0$ in energy, i.e., when the wave-function mixing between states in band $n\neq 1$ and $0$ are small, the summation can be restricted to bands $0,1$ and the two-band approximation still holds. \textcolor{black}{To assess the role of remote bands, we consider a simple three-band setting in which a low-energy subspace is coupled to a distant band separated by an energy scale $\Delta \gg |\varepsilon_0-\varepsilon_1|$. Using $\mathcal{A}_{nm} = \frac{v_{nm}}{i(\varepsilon_n-\varepsilon_m)}$, inter-band processes involving the remote band necessarily enter through virtual transitions. The resulting correction to the Berry connection polarizability is therefore parametrically suppressed as $\delta G \sim v^2/\Delta^3$, demonstrating that remote-band contributions are sub-leading in the well-separated limit. This confirms the quantitative validity of the two-band description when $\Delta$ is large. We further note that this behavior is consistent with model studies of quantum metric dipole nonlinear responses. For example, in topological antiferromagnetic systems, it has been shown that the dominant contribution arises from Dirac surface states, while remote bands contribute only weakly~\cite{gao2023quantum}.}

Let us now consider the case of two bands denoted as band 0 and band 1, and explicitly illustrate the role of quantum metric dipole in nonlinear thermoelectric effects. Let's further replace the energy difference $\varepsilon_0-\varepsilon_1$ by some typical energy splitting $\bar{\varepsilon}$ which depends approximately linearly on the chemical potential $\mu$.

The quantum metric dipole for a band is defined as \cite{gao2023quantum}:
\begin{equation}
     D_{\text{Metric},abc}(\varepsilon_F)=\hbar\int [dk](g_{ab}v_c-g_{cb}v_a)\delta(\varepsilon-\varepsilon_F).
\end{equation}
Then the intrinsic nonlinear thermoelectric responses can be expressed as:
\begin{align}\label{eq:qm-dipole}
    &\hat{M}^{112}_{abc}=-\frac{e^2}{\hbar T}\int d\varepsilon \frac{(\varepsilon-\mu)}{\bar{\varepsilon}}(\partial_{\varepsilon}f)D_{\text{Metric},abc}(\varepsilon),\\
    &\hat{M}^{212}_{abc}=-\frac{e}{\hbar T}\int d\varepsilon \frac{(\varepsilon-\mu)^2}{\bar{\varepsilon}}(\partial_{\varepsilon}f)D_{\text{Metric},abc}(\varepsilon).
\end{align}

\textcolor{black}{From Sommerfeld expansion we can derive that: 
\begin{align}\label{eq:M_two_band_Mott}
    \hat{M}_{abc}^{112}(\mu)=\frac{e^2}{\hbar}\frac{\pi^2}{3}(k_B^2T)\frac{d}{d\mu}\left[\frac{D_{\text{Metric},abc}(\mu)}{\bar{\varepsilon}(\mu)} \right]|_{\mu=\varepsilon_F}+\mathcal{O}(T^3),
\end{align}
and 
\begin{align}\label{eq:M_two_band_WF}
    \hat{M}_{abc}^{212}(\mu)=\frac{e}{\hbar}\frac{\pi^2}{3}(k_B^2T)\frac{D_{\text{Metric},abc}(\varepsilon_F)}{\bar{\varepsilon}(\varepsilon_F)}+\mathcal{O}(T^3).
\end{align}}
Therefore, at low temperatures, the quantum metric dipole and its derivative with respect to the chemical potential directly govern the intrinsic nonlinear thermoelectric effects. Equivalently, the quantum metric dipole can be extracted through nonlinear thermoelectric measurements in this regime.

\section{A quantized nonlinear thermoelectric effect}\label{sec:quantized_response}
One remarkable consequence of our results is a quantized nonlinear thermoelectric effect, which we describe in this section.
First, the trace of the Berry curvature dipole tensor is a constant that reflects the total chirality $\mathcal{S}_{\text{tot}}$ (following the notation of Ref.~\onlinecite{matsyshyn2019nonlinear}) of Weyl nodes below the Fermi energy and can be represented as:
\begin{equation}\label{eq:trace_BCD}
    \text{Tr}(D_{\text{Berry}})=\frac{\mathcal{S}_{\text{tot}}}{4\pi^2}.
\end{equation}

When inversion symmetries and mirror symmetries are broken (i.e., in chiral structures), Weyl nodes with opposite chiralities are in general at different energies and hence $\mathcal{S}_{\text{tot}}$ can be non-zero when the chemical potential lies within certain energy windows. 

Our crucial observation is that the trace of the Berry curvature dipole can be obtained by performing a cyclic summation over the thermoelectric coefficients $\hat{L}^{212}_{abc}$ where $a,b,c$ cycle through the permutations of spatial directions $x,y,z$:
\begin{equation}\label{eq:L212_cyclicsum}
    \begin{split}
    &\sum\limits_{\text{cyc. sum}}\hat{L}^{212}\equiv\hat{L}_{xyz}^{212}+\hat{L}_{yzx}^{212}+\hat{L}_{zxy}^{212}\\
    &= \frac{e\tau}{\hbar^2}\frac{i\omega\tau}{i\omega\tau+1}\int d\varepsilon \frac{(\varepsilon-\mu)^2}{T}(\partial_{\varepsilon}f)\text{Tr}(D_{\text{Berry}}(\varepsilon)).       
    \end{split}
\end{equation}

Now consider a system in an AC electric field with $\omega\tau\gg 1$ at low temperature (e.g., under THz light). In this limit the above expression can be simplified to the following form:
\begin{align}\label{eq:thermal_chirality}
    \sum\limits_{\text{cyc. sum}}\hat{L}^{212} \simeq -\frac{e^3}{h^2}L_0T\tau\cdot \mathcal{S}_{\text{tot}},
\end{align}
where we have used the Sommerfeld expansion Eq.\eqref{eq:L_WiedemannFranzlaw} and the definition Eq.\eqref{eq:trace_BCD}, we have approximated $\frac{i\omega\tau}{i\omega\tau+1}\approx 1$, and $L_0\equiv \frac{\pi^2}{3}(\frac{k_B}{e})^2$ is the Lorenz number.

Eq.\eqref{eq:thermal_chirality} is the central prediction of this section. Put another way, it predicts the following quantization law: 
\begin{align}\label{eq:chirality_quantization}
& \left(\sum\limits_{\text{cyc. sum}}\hat{L}^{212} \right)/ \left(-\frac{e^3}{h^2}L_0T\tau \right) \simeq \mathcal{S}_{\text{tot}},
\end{align}
with $\mathcal{S}_{\text{tot}}$ an integer.

Several comments about this result are in order. 
First, the experiments can be done as the combination of three separate experiments. The only non-universal constant on the RHS of Eq.~\eqref{eq:chirality_quantization} is the relaxation time $\tau$, which can be extracted, for example, by ultra-fast pump-probe experiments. An alternative approach would be to do the measurement of $\sum\limits_{\text{cyc. sum}} \hat{L}^{212}$ with electric fields at different frequencies, thereby extracting both the value of $\tau$ and the part $(-\frac{e^3}{h^2}L_0T\cdot \mathcal{S}_{tot})$ by fitting the following form as a function of $\omega$:
\begin{equation}\label{eq:simplified_L212}
   \sum\limits_{\text{cyc. sum}}\hat{L}^{212}(\omega)\simeq (-\frac{e^3}{h^2}L_0T\cdot \mathcal{S}_{tot})\cdot \frac{i\omega \tau^2}{i\omega\tau+1}.
\end{equation}
\textcolor{black}{Specifically, by measuring the response at two distinct frequencies, $\omega_1$ and $\omega_2$, and taking the ratio of these signals, one obtains an equation involving $\tau$ as the only unknown parameter according to Eq.~\eqref{eq:simplified_L212}. The relaxation time can then be determined directly and used to isolate the universal topological contribution to the response. }

Second, a key advantage of the nonlinear thermoelectric effect over the nonlinear Hall effect lies in the fact that the temperature gradient is applied at zero frequency, while the electric field can have a finite frequency. This distinction allows us to access the trace of the Berry curvature dipole at large $\omega$. In contrast, the nonlinear Hall effect behaves quite differently. Starting from the expression for the BCD-induced nonlinear Hall response \cite{sodemann2015quantum}: $\sigma_{\text{BCD},abc}=-\epsilon_{adc}\frac{e^3\tau}{2(1+i\omega\tau)}D_{\text{Berry},bd}$ and applying the symmetrization procedure discussed in Ref.\cite{matsyshyn2019nonlinear}—namely ($b \leftrightarrow c$, $\omega \leftrightarrow -\omega$)—we obtain: $\sigma_{\text{BCD},xyz}+\sigma_{\text{BCD},yzx}+\sigma_{\text{BCD},zxy}=\frac{-e^3\tau}{2(1+i\omega\tau)}\text{Tr}(D_{\text{Berry}})+\frac{e^3\tau}{2(1-i\omega\tau)}\text{Tr}(D_{\text{Berry}})=\frac{ie^3\omega\tau^2}{1+\omega^2\tau^2}\text{Tr}(D_{\text{Berry}})$. This expression decays as $1/\omega$ at large $\omega$ and becomes appreciable only in a narrow window around $\omega \sim 1/\tau$, making experimental detection challenging. We also briefly comment on a recent proposal, Ref.~\onlinecite{peshcherenko2024quantized}, which employs a cyclic experimental protocol to probe quantized nonlinear Hall responses. Their setup relies on the difference in scattering times between Weyl nodes of opposite chirality, leading to a low-frequency response proportional to $\tau_{+} - \tau_{-}$. At high frequencies, the nonlinear Hall response still decays as $1/\omega$ following our discussion above. For the material CoSi studied in Ref.\onlinecite{peshcherenko2024quantized}, $\mathcal{S}_{\text{tot}} = 0$, and thus Eq.\eqref{eq:chirality_quantization} appears to yield zero. However, due to the imbalance between $\tau_{+}$ and $\tau_{-}$, the nonlinear thermoelectric response described above still survives and yields a value $\left(\sum\limits_{\text{cyc. sum}}\hat{L}^{212}\right) / \left[-\frac{e^3}{h^2}L_0T(\tau_{+} - \tau_{-})\right] \simeq C$, where $C$ denotes the total number of pairs of Weyl nodes below the Fermi energy.

Third, unlike the seminal prediction of the quantized circular photogalvanic effects (CPGE)\cite{de2017quantized,rees2020helicity,ni2021giant,flicker2018chiral}, our result remains robust despite variations in band geometry and interactions, and holds well within an extensive temperature range. The quantized CPGE relies on a low-energy two-band approximation and therefore it receives corrections from higher bands. CPGE also corresponds to real optical absorption processes, and therefore the quantization law is modified when interactions are present\cite{avdoshkin2020interactions}. Our nonlinear thermoelectric response, in contrast, is a Fermi liquid property that involves only the quasiparticles at the Fermi surface at low temperatures. Consequently, it is expected to remain quantized regardless of band dispersions and interactions. \textcolor{black}{More explicitly, following the prescription of Refs.~\onlinecite{haldane2004berry,haldane2004ykis,wang2026interaction}, the fact that the inverse of the exact Green's function is strictly Hermitian at the Fermi surface allows the quasiparticle Bloch state to be defined as its eigenvector with a zero eigenvalue. Consequently, the Berry connection and Berry curvature are well-defined on the Fermi surface even for an interacting system, resulting in the robustness of the quantized response. We defer a more detailed discussion to Appendix.~\ref{appendix:robustness}.}.

Now let's consider the temperature dependence. Notice that $\mathcal{S}_{\text{tot}}$ is a step function and only changes value when the Fermi energy sweeps across a Weyl node. Assuming that the closest Weyl node is at an energy $\pm \Delta$ relative to the chemical potential, the weighting function in Eq.\eqref{eq:L212_cyclicsum} can only take values beyond that energy scale when $k_B T \gtrsim \Delta$. We therefore expect the quantization law Eq.\eqref{eq:chirality_quantization} to hold well when $k_BT \ll \Delta$. Given that $\Delta$ is typically on the order of several tens of meV, the temperature range in which the quantization law holds well in practice can be quite broad.

Ideal candidate materials for observing the quantized nonlinear thermoelectric effect are those with a large energy difference between Weyl nodes of opposite chirality. For example, the chiral Weyl semimetal $\text{SrSi}_2$ has an energy offset $\sim 0.1$\,eV between double Weyl nodes\cite{huang2016new}, CoSi has an energy off-set $\sim 0.2$\,eV \cite{ni2021giant,takane2019observation}, and RhSi has an energy off-set $\sim 0.65$\,eV \cite{rees2020helicity}.

\begin{figure}[htb]
    \centering
    \includegraphics[width=0.45\textwidth]{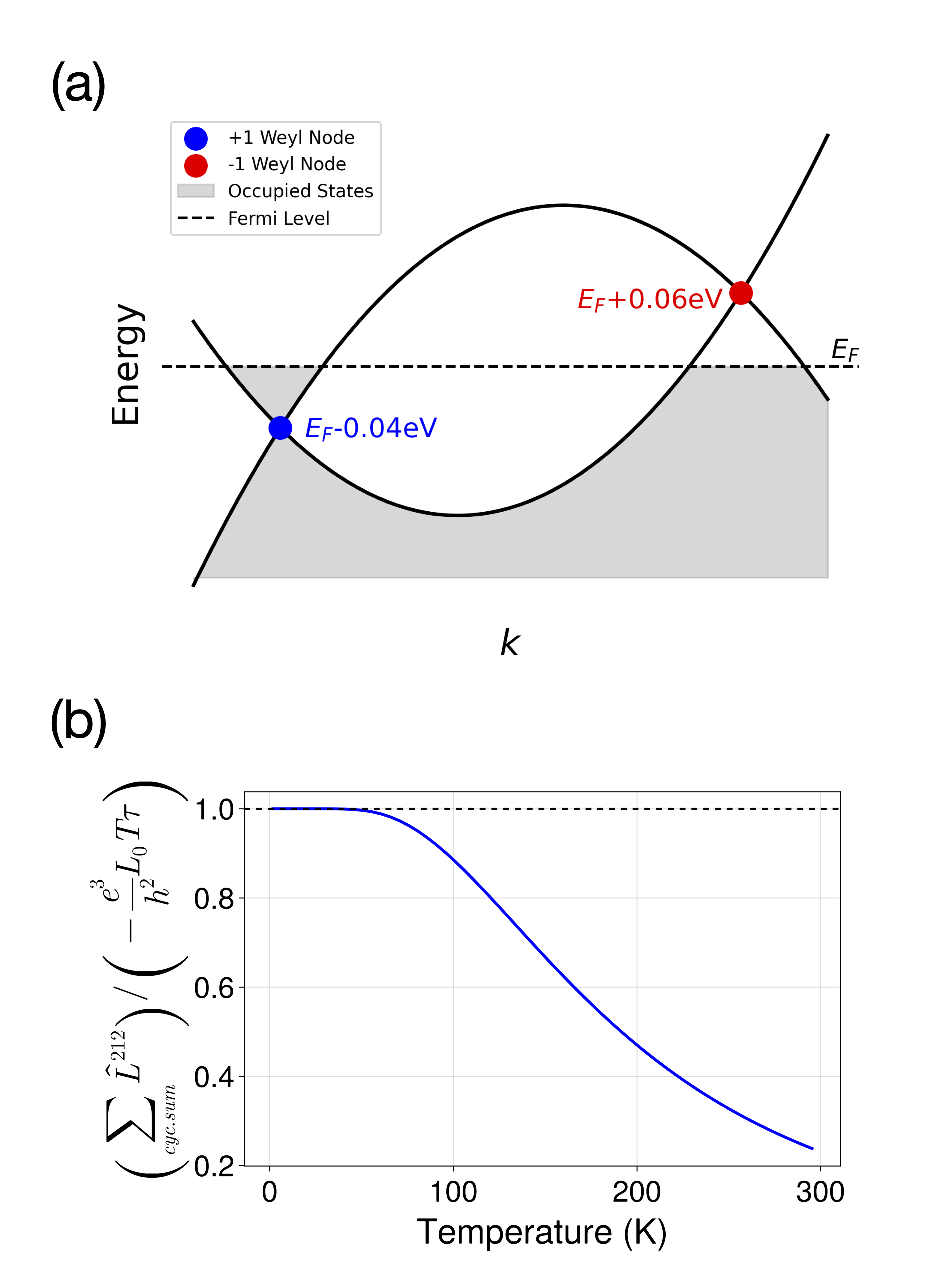}
    \caption{Nonlinear thermoelectric responses dominated by the trace of the Berry curvature dipole. (a) A schematic picture of the bands. We consider a pair of Weyl nodes with opposite chiralities. The energy of the ``+1" Weyl node is lower than that of the ``-1" Weyl node with an energy off-set of $0.1eV$. The Fermi energy is $0.04eV$ above the ``+1" node. (b) The temperature dependence of $\sum\limits_{\text{cyc. sum}}\hat{L}^{212}\equiv \hat{L}^{212}_{xyz}+\hat{L}^{212}_{yzx}+\hat{L}^{212}_{zxy}$. At temperatures below $\sim 50K$, the Sommerfeld expansion is highly accurate, from which we can extract the quantized value of the total chirality $\mathcal{S}_{\text{tot}}$ below the Fermi energy.  } 
    \label{fig2}
\end{figure}

We now demonstrate the temperature dependence of the quantization law Eq.~\eqref{eq:chirality_quantization} by a simple example calculation that considers a single pair of Weyl nodes with opposite chiralities for which the energy of the ``+1" Weyl node is lower than the energy of the ``-1" Weyl node by $0.1$\,eV. We take the Fermi energy $\varepsilon_F$ to be $0.04$\,eV above the ``+1" Weyl node and we calculate $\sum\limits_{\text{cyc. sum}}\hat{L}^{212}$ using the integral expression of Eq.~\eqref{eq:L212_cyclicsum}. The chemical potential $\mu$ depends only very weakly on temperature at $k_B T \ll \varepsilon_F$, so we assume a constant $\mu=0.04$\,eV in our calculation. Thus we have $\text{Tr}(D_{\text{Berry}}(\varepsilon-\varepsilon_F))=1/(4\pi^2)$ for $-0.04\,\text{eV}<\varepsilon-\varepsilon_F<0.06\,\text{eV}$ and zero elsewhere. The results are displayed in Fig.~\ref{fig2}, and a quantization plateau of $\left(\sum\limits_{\text{cyc. sum}}\hat{L}^{212}\right)/\left(-\frac{e^3}{h^2}L_0T\tau\right)\approx 1$ is clearly observed for temperatures below $\approx 50K$.

\textcolor{black}{We conclude this section by discussing the experimental conditions on $\omega$ and $T$ for observing such quantized responses. For a system with a relaxation time of approximately $\tau \sim 10\,$ps, the condition $\omega\tau \gg 1$ can be readily achieved using terahertz-frequency light. Furthermore, for the candidate materials SrSi$_2$, CoSi and RhSi, the characteristic energy offset is $\Delta \gtrsim 0.1\,$eV. We note that for $T \lesssim 100\,$K, the condition $k_B T \ll \Delta$ is well satisfied, and the next-to-leading correction in the Sommerfeld expansion of Eq.~\eqref{eq:L212_cyclicsum} is small, as confirmed in Fig.~\ref{fig2}. }

\section{Model calculations}
\label{sec:model}
In this section, we use two models to illustrate nonlinear thermoelectric effects. The model calculations also serve as a guide for optimizing signals based on doping levels and temperatures.

\subsection{Weyl node with band-bending for BCD-induced nonlinear thermoelectric effects}\label{subsec:modelB}

\begin{figure*}
    \centering
    \includegraphics[width=1\textwidth]{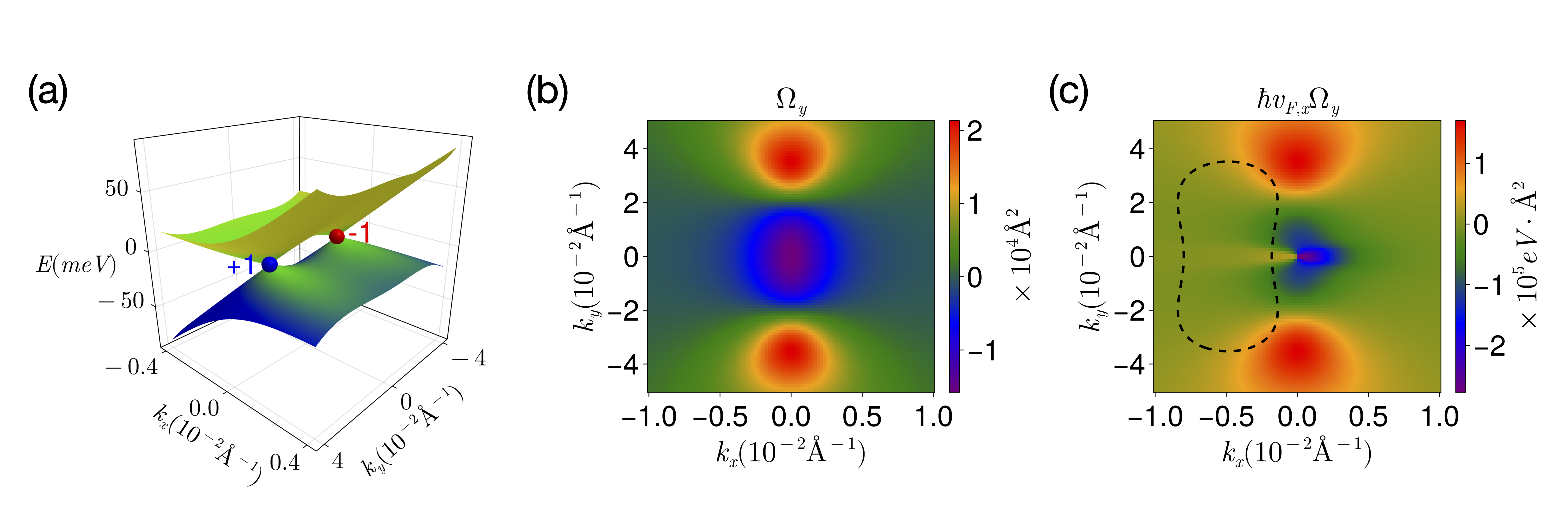}
    \caption{Plot of the dispersion relation, Berry curvature and Berry curvature dipole of the model Hamiltonian defined in Eq.~\eqref{eq:Weyl_bending}. (a) Dispersion relation of the Hamiltonian in the $k_z=0$ plane, which consists of two Weyl nodes with opposite chiralities that are separated along the $k_y$ direction. (b) The $y$ component of Berry curvature, $\Omega_{y}$, in the $k_x$-$k_y$ plane with $k_z=0.3$ \AA$^{-1}$. Notice that the Berry curvature distribution is centered around the two Weyl nodes. (c) The distribution of the quantity $\hbar v_{F,x}\Omega_y$ in the $k_x$-$k_y$ plane with $k_z=0.3$ \AA$^{-1}$. Integrating this quantity over the Fermi surface gives the Berry curvature dipole $D_{xy}$. We have also plotted the contour describing the Fermi surface with $\varepsilon_F=22$\,meV.
    }
\label{fig:dispersion&bcd}
\end{figure*}

The necessary ingredient to capture the anti-symmetric part of the BCD is the introduction of quadratic terms in the dispersion relation. Therefore we consider the following model Hamiltonian, which describes two Weyl nodes of opposite chirality\cite{facio2018strongly}:
\begin{align}\label{eq:Weyl_bending}
    H=\hbar v_xk_x\sigma_x+\frac{\lambda-(\hbar k_y)^2}{2m}\sigma_y+\hbar v_zk_z\sigma_z+\hbar u_xk_x\sigma_0.
\end{align}
For $\lambda>0$, there are two Weyl nodes separated by a distance of $2\sqrt{\lambda}/\hbar$. (The velocity near the Weyl point in the $y$ direction is $v_y=\sqrt{|\lambda|}/m$.)  For negative $\lambda$ there is a band gap between the conduction and valence bands equal to $\Delta=\frac{|\lambda|}{m}$. Numerical plots below use the following parameter values: $\hbar v_y=1$\,eV\AA, $v_x=v_z=10v_y,u_x=8v_y, \frac{\lambda}{2m}=10$\,meV, and $\tau=10$\,ps. See Fig.~\ref{fig:dispersion&bcd} for an illustration of the dispersion relation, Berry curvature, and Berry curvature dipole of this model Hamiltonian.
We take $\varepsilon_F=7$\,meV and study the temperature dependence of various thermoelectric responses.

\textcolor{black}{A note on the choice of parameters: this model describes a pair of Weyl nodes with finite tilt $u_x=0.8v_x$, where the Fermi velocities are $v_x=v_z\sim 10^6$\,m/s, which are typical values for Weyl semimetals. The relaxation time corresponds to that of a relatively clean system, while $v_y$ is taken to be suppressed due to the proximity of the two Weyl nodes. The nodes are separated by $2\sqrt{\lambda}/\hbar\sim 0.04 $\AA$^{-1}$, consistent with a scenario where two Weyl nodes emerge from the splitting of a Dirac semimetal under weak inversion-symmetry breaking.}

\begin{figure*}
    \centering
    \includegraphics[width=1\textwidth]{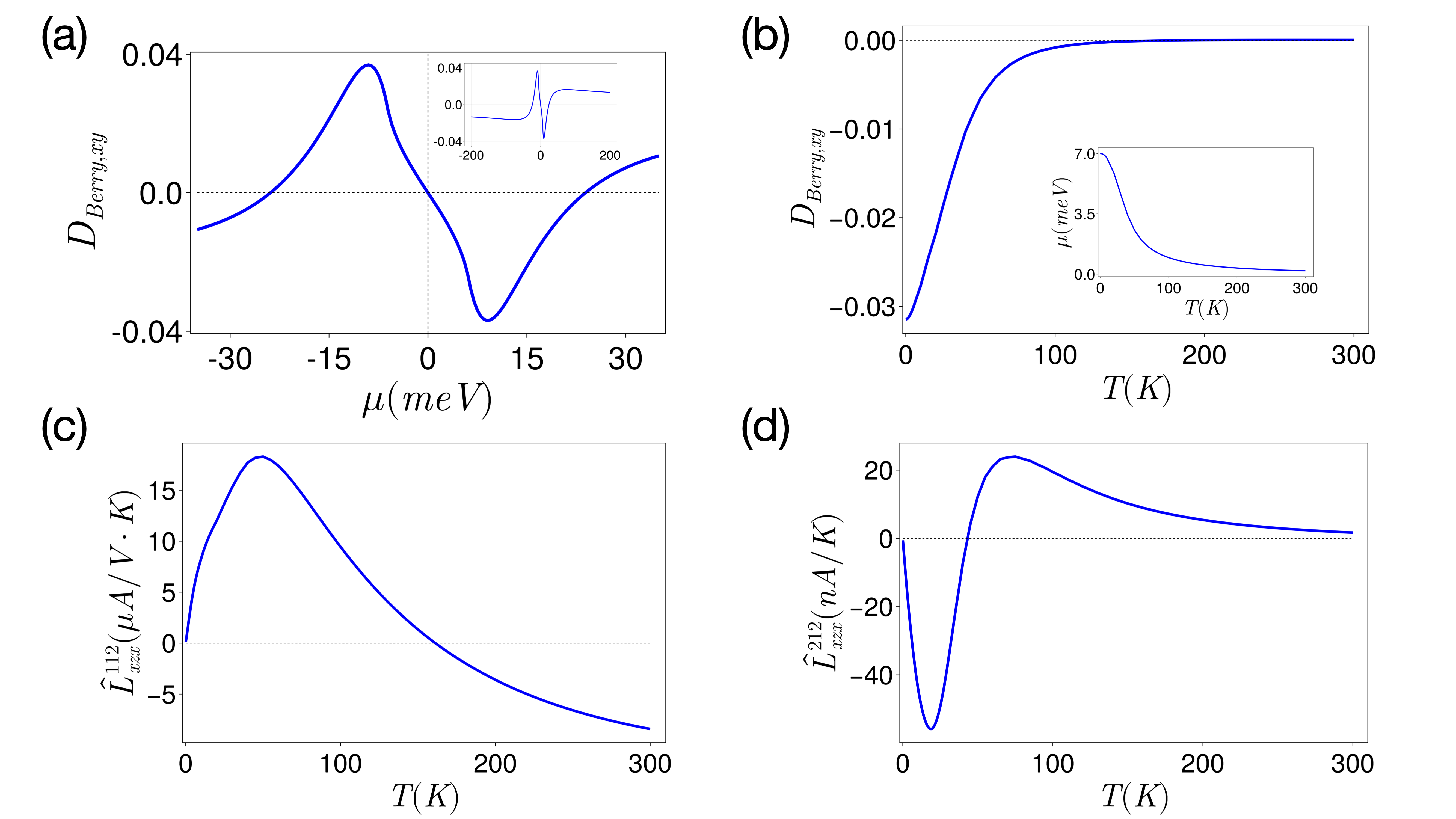}
    \caption{The Berry curvature dipole contribution $D_{\text{Berry},xy}$ to the nonlinear electric/thermoelectric responses for the model Hamiltonian shown in Fig.~\ref{fig:dispersion&bcd} (a) The $\mu$ dependence of Berry curvature dipole $D_{\text{Berry},xy}$ at $T = 0$. The inset shows $D_{\text{Berry},xy}$ over a broader range of chemical potential. (b) The temperature dependence of $D_{\text{Berry},xy}$ with a fixed $\varepsilon_F=7$\,meV. The inset shows the temperature dependence of $\mu$ assuming a fixed electron density. (c-d) The temperature dependence of $\hat{L}^{112}_{xzx}$ and $\hat{L}^{212}_{xzx}$ with a fixed $\varepsilon_F=7$\,meV. The $T$ dependence of $\mu$ is the same as that in the inset of (b).}
\label{fig:dxy}
\end{figure*}

In the following we focus on the anti-symmetric part $(D_{\text{Berry},xy}-D_{\text{Berry},yx})/2$ of the BCD. This is because, for materials such as TaAs in transition metal monopinicide family, the crystalline point group $C_{4v}$ constraints the BCD tensor such that the only non-zero elements are $D_{\text{Berry},xy}=-D_{\text{Berry},yx}$ (see Table.~\ref{tab:pseudotensor_symmetry}). As a result, only the anti-symmetric component of a single Weyl node pair contributes to the total $D_{\text{Berry},xy}$ after averaging over all symmetry-related Weyl node pairs. The thermoelectric responses are computed via Eq.\eqref{eq:bcdintegration1}-\eqref{eq:bcdintegration2} and the results are displayed in Fig.~\ref{fig:dxy}. Two remarks are in order. First, at high temperatures, the Berry curvature dipole is averaged over a broad energy window and becomes smeared out. Second, in the low-temperature regime, the Sommerfeld expansions [Eq.\eqref{eq:L_Mottformula}, Eq.\eqref{eq:L_WiedemannFranzlaw}] hold well, and the coefficients $L^{112}$ and $L^{212}$ are expressed in terms of the zero-temperature Berry curvature dipole and its energy derivatives, multiplied by temperature $T$ and universal constants. As a result, they exhibit a linear-in-$T$ enhancement at low temperatures, which is evident in Fig.\ref{fig:dxy}(c,d).

\subsection{Massive Dirac node for QMD-induced nonlinear thermoelectric effects}\label{subsec:modelC}
\begin{figure*}
    \centering
    \includegraphics[width=1\textwidth]{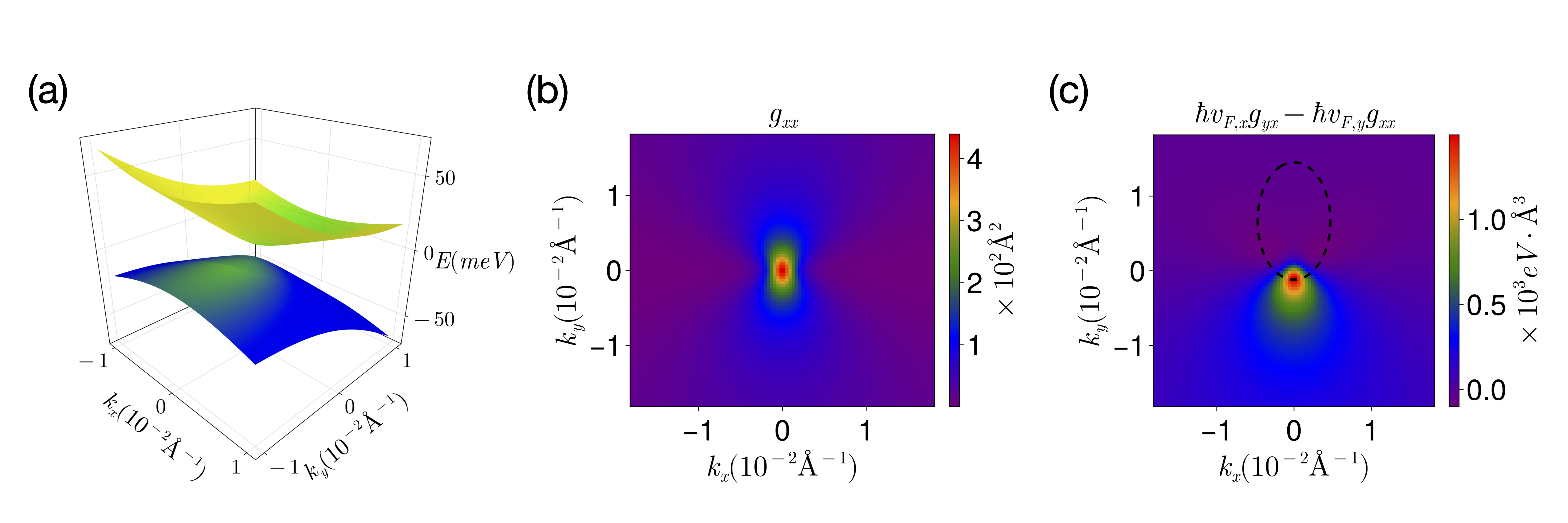}
    \caption{Plot of the dispersion relation, quantum metric and quantum metric dipole of the model Hamiltonian defined in Eq.~\eqref{massive_dirac}. (a) Dispersion relation of the Hamiltonian in the $k_z=0$ plane. (b) Distribution of the quantum metric $g_{xx}$ in the $k_x-k_y$ plane with $k_z=0$. Notice that the quantum metric is centered around the massive Dirac point. The value of $g_{xy}$ is an order of magnitude smaller than $g_{xx}$ and is therefore omitted here for clarity. (c) Distribution of the quantum metric dipole $\hbar (v_xg_{yx}-v_yg_{xx})$ in the $k_x-k_y$ plane with $k_z=0$. We have also plotted the contour describing the Fermi surface with $\varepsilon_F=9$\,meV. The quantum metric dipole is primarily concentrated around $k_x=k_y=0$. 
    }
\label{fig:dispersion&qmd}
\end{figure*}

We now consider a massive Dirac node respecting $\mathcal{P}\mathcal{T}$ symmetry following Ref.\cite{wang2021INH,liu2021INH}. Owing to the $\mathcal{P}\mathcal{T}$ symmetry, the Berry curvature vanishes at every $k$ point; as a result, the Berry curvature dipole is absent, making the quantum metric dipole effect stand out as the dominant contribution to the nonlinear thermoelectric effect. The Hamiltonian is:
\begin{equation}\label{massive_dirac}
H= \hbar(v_xk_x\tau_x\sigma_0+ v_yk_y\tau_y\sigma_y- u_yk_y\tau_0\sigma_0+ v_zk_z\tau_z\sigma_0)+\Delta\tau_y\sigma_z,
\end{equation}
where $\sigma$ and $\tau$ are two sets of Pauli matrices acting on the spin and orbital degrees of freedom, respectively, following the notation of Ref.~\cite{wang2021INH}. The $\mathcal{P}\mathcal{T}$ symmetry is represented by $-i\sigma_y\mathcal{K}$ where $\mathcal{K}$ is the complex conjugation operator taking $i\rightarrow -i$. \textcolor{black}{We take $\Delta = 5\,\mathrm{meV}$, $v_x=v_y=v_z=3\,\mathrm{eV\AA}/\hbar$, and $u_y=0.8v_z$, corresponding to a Fermi velocity $v \sim 10^5\,\mathrm{m/s}$ and a small gap scale typical of narrow-gap Dirac semimetals. The tilt parameter $u_y / v_z = 0.8$ describes a moderately tilted Dirac dispersion.} In determining the temperature dependence of the chemical potential and hence the nonlinear thermoelectric responses, we have chosen $\varepsilon_F=9$\,meV and assume the total electron density is fixed. See Fig.\ref{fig:dispersion&qmd} for an illustration of the dispersion relation and the distribution of quantum metric and quantum metric dipole in the Brillouin zone.

\begin{figure*}
    \centering
    \includegraphics[width=1\textwidth]{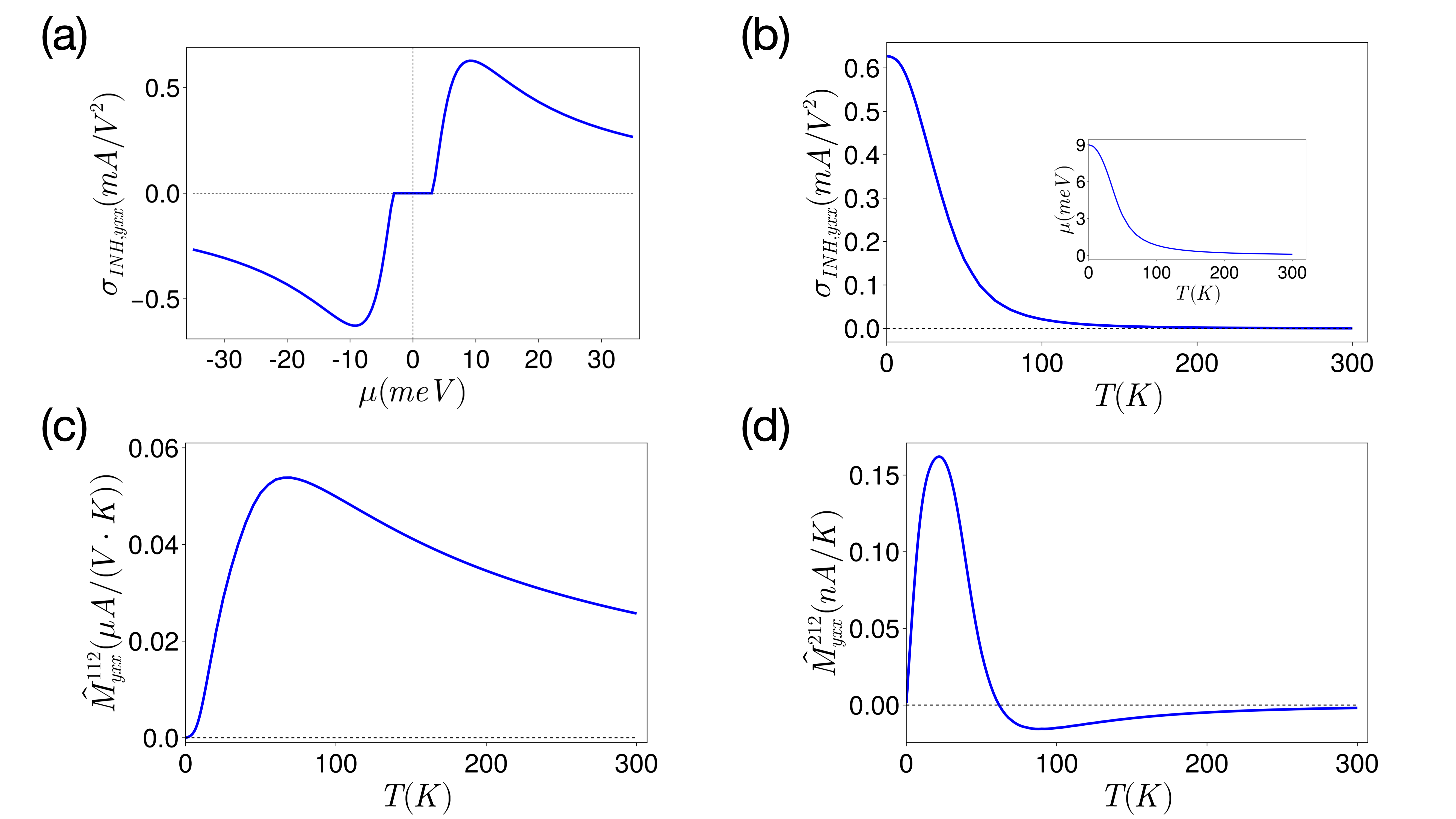}
    \caption{The intrinsic nonlinear electric/thermoelectric responses for the model Hamiltonian depicted in Fig.~\ref{fig:dispersion&qmd}. (a) The $\mu$ dependence of the intrinsic nonlinear Hall response $\sigma_{{\text{INH}},yxx}$ at $T=0$. (b) The temperature dependence of $\sigma_{\text{INH},yxx}$ with a fixed $\varepsilon_F=9$\,meV. The inset shows the temperature dependence of the chemical potential $\mu$ assuming a fixed electron density. (c-d) The temperature dependence of $\hat{M}_{yxx}^{112}$ and $\hat{M}_{yxx}^{212}$ with a fixed  $\varepsilon_F=9$\,meV. The $T$ dependence of $\mu$ is the same as that in the inset of (b). This example uses $m=5$\,meV, $v_x=v_y=v_z=3$\,eV\AA$/\hbar$, $u_y=0.8v_z$ as parameters for the Hamiltonian in Eq.\eqref{massive_dirac}.  }
    \label{fig:QMD}
\end{figure*}
With these parameters we have computed the intrinsic nonlinear thermoelectric effects induced by the quantum metric dipole using Eqs.~\eqref{eq:sigma_INH}-\eqref{eq:QMD_integrations2}. The results are displayed in Fig.~\ref{fig:QMD}.

The behavior of the QMD-induced nonlinear thermoelectric effects discussed here closely parallels that of the BCD-induced nonlinear thermoelectric response discussed in the previous subsection. In particular, the quantum metric dipole is thermally smeared out at high temperatures resulting in a suppression of $\sigma_{{\text{INH}}}$. The responses $\hat{M}^{112}$ and $\hat{M}^{212}$ also exhibit a linear-in-$T$ enhancement at low temperatures, consistent with the Sommerfeld expansions [Eq.\eqref{eq:M_Mott_formula} which relate them to $\sigma_{{\text{INH}}}$ and its energy derivatives multiplied by $T$. In the two-band limit considered here, Eq.\eqref{eq:M_two_band_Mott} and Eq.\eqref{eq:M_two_band_WF} further allow for direct extraction of the quantum metric dipole and its energy derivative.

\section{Experimental consequences}
\label{sec:experiment}
\subsection{Symmetry constraints}
The symmetry properties of $\hat{L}$ and $\hat{M}$ dictate the symmetries of the BCD-induced nonlinear Hall effects and intrinsic nonlinear Hall effects. Both effects require the breaking of inversion symmetry. The Berry curvature dipole can be non-zero in systems respecting $\mathcal{T}$ symmetry, but it is zero in systems respecting $\mathcal{P}\mathcal{T}$ symmetry. Its symmetry transformation rule can be found in Ref.~\onlinecite{sodemann2015quantum,zhang2023symmetry}. \textcolor{black}{For completeness we list the non-zero components of $D_{\text{Berry},ab}$ for all the point groups (PGs) in Table.~\ref{tab:pseudotensor_symmetry}.} The Quantum metric dipole is zero in systems respecting $\mathcal{T}$ symmetry but can be nonzero in systems respecting $\mathcal{P}\mathcal{T}$ symmetry. A complete symmetry analysis of the intrinsic nonlinear Hall effects can be found in Table I of Ref.~\onlinecite{wang2021INH} and applies to $\hat{M}^{112}$ and $\hat{M}^{212}$.

\begin{table}[h]
\centering
\begin{tabular}{|c|c|}
\hline
\textbf{NCS PGs} & \textbf{Matrix form of $D_{\text{Berry},ab}$} \\
\hline
$C_1$
&
General form
\\
\hline
$C_2$ & 
$\begin{pmatrix}
D_{xx} & D_{xy} & 0 \\
D_{yx} & D_{yy} & 0 \\
0 & 0 & D_{zz}
\end{pmatrix}$ 
\\  
\hline
$C_3,C_4,C_6$ & 
$\begin{pmatrix}
D_{xx} & D_{xy} & 0 \\
-D_{xy} & D_{xx} & 0 \\
0 & 0 & D_{zz}
\end{pmatrix}$ 
\\  
\hline
$C_{s}$ & 
$\begin{pmatrix}
0 & D_{xy} & D_{xz} \\
D_{yx} & 0 & 0 \\
D_{zx} & 0 & 0
\end{pmatrix}$ 
\\  
\hline
$C_{2v}$ & 
$\begin{pmatrix}
0 & D_{xy} & 0 \\
D_{yx} & 0 & 0 \\
0 & 0 & 0
\end{pmatrix}$ 
\\  
\hline
$C_{3v}, C_{4v}, C_{6v}$
&
$\begin{pmatrix}
0 & D_{xy} & 0 \\
-D_{xy} & 0 & 0 \\
0 & 0 & 0
\end{pmatrix}$
\\
\hline
$D_2$ & 
$\begin{pmatrix}
D_{xx} & 0 & 0 \\
0 & D_{yy} & 0 \\
0 & 0 & D_{zz}
\end{pmatrix}$ 
\\  
\hline
$D_3,D_4,D_6$ & 
$\begin{pmatrix}
D_{xx} & 0 & 0 \\
0 & D_{xx} & 0 \\
0 & 0 & D_{zz}
\end{pmatrix}$ 
\\  
\hline
$O,T$
&
$D_{\text{Berry},ab}\propto \delta_{ab}$
\\
\hline
$S_4$
&
$\begin{pmatrix}
D_{xx} & D_{xy} & 0 \\
D_{xy} & -D_{xx} & 0 \\
0 & 0 & 0
\end{pmatrix}
$
\\
\hline
\end{tabular}
\caption{
Symmetry-allowed forms of the BCD tensor $D_{\text{Berry},ab}$, which transforms as $D'=\det(U)UDU^T$ under a point group operation $U$. Only non-centrosymmetric systems (NCS) permit a nonzero BCD. The coordinate system is chosen such that, whenever present, the principal rotation axis is aligned with the $z$-axis and the mirror plane $m_v$ coincides with the $yz$ plane. For the cubic groups $O$ and $T$, $D_{\text{Berry},ab}\propto \delta_{ab}$. Point groups not listed do not admit a non-zero BCD.
}
\label{tab:pseudotensor_symmetry}
\end{table}

The Levi-Civita symbol in the definition of the $\hat{L}$ tensors implies constraints on their non-zero components. Therefore we have listed all possible components of $\hat{L}^{112}$ and $\hat{L}^{212}$, as well as the Berry curvature dipoles related to them via Eq.~\eqref{eq:bcdintegration1}/\eqref{eq:bcdintegration2}, in Tables \ref{tab:thermoelectric112} and \ref{tab:thermoelectric212}, respectively. The coefficients $\hat{M}^{112}_{abc}$ and $\hat{M}^{212}_{abc}$ are anti-symmetric with respect to exchange of $a,c$ and are otherwise unconstrained.

\begin{table}[]
    \centering
    \begin{tabular}{|c|c|c|c|}
    \hline
Modified & Temperature  & Corresponding &Corresponding \\
 $\sigma$& gradient direction& $\hat{L}^{112}$ & BCDs \\
         \hline
         \multirow{ 3}{*}{$\sigma_{xx}$} & $\nabla T_x$ &$-$&0 \\
 \cline{2-4}           & $\nabla T_y$ &$\hat{L}^{112}_{xxy}$&$D_{xz}$ \\
 \cline{2-4} 
                  & $\nabla T_z$ &$\hat{L}^{112}_{xxz}$&$-D_{xy}$ \\
         \hline

                  \multirow{ 3}{*}{$\sigma_{xy}$} & $\nabla T_x$ &$\hat{L}^{112}_{xyx}$&$-2D_{xz}$ \\
 \cline{2-4} 
& $\nabla T_y$ &$\hat{L}^{112}_{xyy}$&$-D_{yz}$ \\  \cline{2-4} 
& $\nabla T_z$ &$\hat{L}^{112}_{xyz}$&$D_{xx}-D_{zz}$ \\ \hline
 \multirow{ 3}{*}{$\sigma_{xz}$} & $\nabla T_x$ &$\hat{L}^{112}_{xzx}$&$2D_{xy}$ \\ \cline{2-4}   
& $\nabla T_y$ &$\hat{L}^{112}_{xzy}$&$-D_{xx}+D_{yy}$ \\ \cline{2-4} 
 & $\nabla T_z$ &$\hat{L}^{112}_{xzz}$&$D_{zy}$ \\ \hhline{|=|=|=|=|}

 \multirow{ 3}{*}{$\sigma_{yx}$} & $\nabla T_x$ &$\hat{L}^{112}_{yxx}$&$D_{xz}$ \\ \cline{2-4}   
& $\nabla T_y$ &$\hat{L}^{112}_{yxy}$&$2D_{yz}$ \\ \cline{2-4} 
 & $\nabla T_z$ &$\hat{L}^{112}_{yxz}$&$-D_{yy}+D_{zz}$ \\ \hline

  \multirow{ 3}{*}{$\sigma_{yy}$} & $\nabla T_x$ &$\hat{L}^{112}_{yyx}$&$-D_{yz}$ \\ \cline{2-4}   
& $\nabla T_y$ &$-$&$0$ \\ \cline{2-4} 
 & $\nabla T_z$ &$\hat{L}^{112}_{yyz}$&$D_{yx}$ \\ \hline

  \multirow{ 3}{*}{$\sigma_{yz}$} & $\nabla T_x$ &$\hat{L}^{112}_{yzx}$&$D_{yy}-D_{xx}$ \\ \cline{2-4}   
& $\nabla T_y$ &$\hat{L}^{112}_{yzy}$&$-2D_{yx}$ \\ \cline{2-4} 
 & $\nabla T_z$ &$\hat{L}^{112}_{yzz}$&$ -D_{zx}$ \\ \hhline{|=|=|=|=|}

  \multirow{ 3}{*}{$\sigma_{zx}$} & $\nabla T_x$ &$\hat{L}^{112}_{zxx}$&$-D_{xy}$ \\ \cline{2-4}  
& $\nabla T_y$ &$\hat{L}^{112}_{zxy}$&$D_{zz}-D_{yy}$ \\ \cline{2-4} 
 & $\nabla T_z$ &$\hat{L}^{112}_{zxz}$&$-2D_{zy}$ \\ \hline

  \multirow{ 3}{*}{$\sigma_{zy}$} & $\nabla T_x$ &$\hat{L}^{112}_{zyx}$&$D_{xx}-D_{zz}$ \\ \cline{2-4}   
& $\nabla T_y$ &$\hat{L}^{112}_{zyy}$&$D_{yx}$ \\ \cline{2-4} 
 & $\nabla T_z$ &$\hat{L}^{112}_{zyz}$&$2D_{zx}$ \\ \hline

  \multirow{ 3}{*}{$\sigma_{zz}$} & $\nabla T_x$ &$\hat{L}^{112}_{zzx}$&$D_{zy}$ \\ \cline{2-4}   
& $\nabla T_y$ &$\hat{L}^{112}_{zzy}$&$-D_{zx}$ \\ \cline{2-4} 
 & $\nabla T_z$ &$-$&$0$ \\ \hline
    \end{tabular}
    \caption{Temperature-gradient-induced electric responses. For a DC electric field, the nonlinear response can be obtained by integrating $-e\tau (\epsilon_{bcd}D_{\text{Berry},ad}+\epsilon_{abd}D_{\text{Berry},cd})$ according to Eq.~\eqref{eq:bcdintegration1}. We have listed the BCDs that are integrated in the last column. At low-temperatures the response is proportional to derivatives of Berry curvature dipoles according to Eq.~\eqref{eq:L_Mottformula}.}
    \label{tab:thermoelectric112}
\end{table}

\begin{table}[]
    \centering
    \begin{tabular}{|c|c|c|c|}
    \hline
         Modified & Electric&Corresponding&Corresponding \\
         $\kappa$ & field direction&$\hat{L}^{212}$& BCDs \\

         \hline
         \multirow{ 3}{*}{$\kappa_{xx}$} & $E_x$&$-$ &0 \\
 \cline{2-4}           & $E_y$ &$\hat{L}^{212}_{xyx}$&$-2D_{xz}$ \\
 \cline{2-4} 
                  & $E_z$ &$\hat{L}^{212}_{xzx}$&$2D_{xy}$ \\
         \hline

                  \multirow{ 3}{*}{$\kappa_{xy}$} & $E_x$ &$\hat{L}^{212}_{xxy}$&$D_{xz}$ \\
 \cline{2-4} 
& $E_y$ &$\hat{L}^{212}_{xyy}$&$-D_{yz}$ \\  \cline{2-4} 
& $E_z$ &$\hat{L}^{212}_{xzy}$&$-D_{xx}+D_{yy}$ \\ \hline

 \multirow{ 3}{*}{$\kappa_{xz}$} & $E_x$ &$\hat{L}^{212}_{xxz}$&$-D_{xy}$ \\ \cline{2-4}   
& $E_y$ &$\hat{L}^{212}_{xyz}$&$D_{xx}-D_{zz}$ \\ \cline{2-4} 
 & $E_z$ &$\hat{L}^{212}_{xzz}$&$D_{zy}$ \\ \hhline{|=|=|=|=|}

 \multirow{ 3}{*}{$\kappa_{yx}$} & $E_x$ &$\hat{L}^{212}_{yxx}$&$D_{xz}$ \\ \cline{2-4}   
& $E_y$ &$\hat{L}^{212}_{yyx}$&$-D_{yz}$ \\ \cline{2-4} 
 & $E_z$ &$\hat{L}^{212}_{yzx}$&$D_{yy}-D_{xx}$ \\ \hline

  \multirow{ 3}{*}{$\kappa_{yy}$} & $E_x$ &$\hat{L}^{212}_{yxy}$&$2D_{yz}$ \\ \cline{2-4}   
& $E_y$ &$-$&$0$ \\ \cline{2-4} 
 & $E_z$ &$\hat{L}^{212}_{yzy}$&$-2D_{yx}$ \\ \hline

  \multirow{ 3}{*}{$\kappa_{yz}$} & $E_x$ &$\hat{L}^{212}_{yxz}$&$-D_{yy}+D_{zz}$ \\ \cline{2-4}   
& $E_y$ &$\hat{L}^{212}_{yyz}$&$D_{yx}$ \\ \cline{2-4} 
 & $E_z$ &$\hat{L}^{212}_{yzz}$&$ -D_{zx}$ \\ \hhline{|=|=|=|=|}

  \multirow{ 3}{*}{$\kappa_{zx}$} & $E_x$ &$\hat{L}^{212}_{zxx}$&$-D_{xy}$ \\ \cline{2-4}  
& $E_y$ &$\hat{L}^{212}_{zyx}$&$-D_{zz}+D_{xx}$ \\ \cline{2-4} 
 & $E_z$ &$\hat{L}^{212}_{zzx}$&$D_{zy}$ \\ \hline

  \multirow{ 3}{*}{$\kappa_{zy}$} & $E_x$ &$\hat{L}^{212}_{zxy}$&$D_{zz}-D_{yy}$ \\ \cline{2-4}   
& $E_y$ &$\hat{L}^{212}_{zyy}$&$D_{yx}$ \\ \cline{2-4} 
 & $E_z$ &$\hat{L}^{212}_{zzy}$&$-D_{zx}$ \\ \hline

  \multirow{ 3}{*}{$\kappa_{zz}$} & $E_x$ &$\hat{L}^{212}_{zxz}$&$-2D_{zy}$ \\ \cline{2-4}   
& $E_y$ &$\hat{L}^{212}_{zyz}$&$2D_{zx}$ \\ \cline{2-4} 
 & $E_z$ &$-$&$0$ \\ \hline
    \end{tabular}
    \caption{Electric-field-induced thermal responses. For a DC field, the response can be obtained by integrating $ (\epsilon_{bcd}D_{\text{Berry},ad}-\epsilon_{abd}D_{\text{Berry},cd})$ according to Eq.\eqref{eq:bcdintegration2}. At low-temperature we can utilize the Sommerfeld expansion and the response is proportional to temperature times the Berry curvature dipoles at $\varepsilon_F$ according to \eqref{eq:L_WiedemannFranzlaw}.}
    \label{tab:thermoelectric212}
\end{table}

Therefore, one can experimentally distinguish the underlying mechanisms of these nonlinear thermoelectric responses by examining their distinct symmetry properties under time-reversal and crystalline symmetries.

\subsection{Experimental set-ups}\label{sec:experimental_meaning}
We can recast the nonlinear transport coefficients in a way that makes their physical meaning transparent. First we notice that the coefficients $\hat{L}^{112},\hat{M}^{112}$ can be viewed as describing the temperature-gradient-induced change of electric conductivity. To be more specific, we have: 
\begin{align}
j_e^a & =\sigma_{ab}E_b+(\hat{L}^{112}_{abc}+\hat{M}^{112}_{abc})E_b\nabla_cT\nonumber\\
&=(\sigma_{ab}+\hat{L}^{112}_{abc}\nabla_cT+\hat{M}^{112}_{abc}\nabla_cT)E_b.
\end{align}
Therefore $\hat{L}^{112}_{abc}$ and $\hat{M}^{112}_{abc}$ can be viewed as the coefficient of linear correction to $\sigma_{ab}$ in the presence of $\nabla_cT$.

Similarly, we have
\begin{align}
j_Q^a & =-\kappa_{ab}\nabla_bT+(\hat{L}^{212}_{acb}+\hat{M}^{212}_{acb})E_c\nabla_bT\nonumber\\
&=(-\kappa_{ab}+\hat{L}^{212}_{acb}E_c+\hat{M}^{212}_{acb}E_c)\nabla_bT,
\end{align}
so that $\hat{L}^{212}$ and $\hat{M}^{212}$ can be viewed as the electric-field induced correction to the thermal conductivity. \textcolor{black}{Note that the above framework yields a response linear in both $E$ and $\nabla T$. From an experimental protocol perspective, this allows one to eliminate contributions proportional to $E_bE_c$ and $(\nabla_bT)(\nabla_cT)$, which are even under sign reversal of either $E$ or $\nabla T$, and thereby isolate the mixed term within this framework.}

\begin{table*}[t]
    \centering
    \begin{tabular}{|c| c| c |p{3cm}| c | c |} 
 \hline
 \bf{Material candidate} & \bf{$T$ symmetry} & \bf{\makecell{Geometric \\ quantity} } & \bf{ \makecell{Measurement \\ setup}} & \bf{\makecell{Values used \\ in estimation}} & \bf{\makecell{Magnitude \\ of response}} \\ [0.5ex] 
 \hline\hline
 \multirow{2}{*}{NbP (TMMP family)} & \multirow{2}{*}{Non-magnetic}& \multirow{2}{*}{$D_{\text{Berry},xy}$}  & $E_z$ induced $\kappa_{xx}$ & $D_{\text{Berry},xy}\sim 20 $ & $4$\,mW/(K\,m)\\  \cline{4-6}
& & & $\nabla_zT$ induced $\sigma_{xx}$ &$\Delta\mu \sim 10$\,meV   & $4$\,S/m \\
 \hline\hline
 \multirow{2}{*}{$\text{MoTe}_2$ (TMDC family)}&\multirow{2}{*}{Non-magnetic}&\multirow{2}{*}{$D_{\text{Berry},xy}$}& $E_z$ induced $\kappa_{xx}$ & $D_{\text{Berry},xy}\sim 1 $ &$0.2$\,mW/(K\,m) \\ \cline{4-6}
 & && $\nabla_zT$ induced $\sigma_{xx}$& $\Delta\mu\sim 50$\,meV & $40$\,mS/m  \\
 \hline\hline
 \multirow{2}{*}{$(\text{Pb}_{1-x}\text{Sn}_x)_{1-y}\text{In}_y\text{Te}$}&\multirow{2}{*}{Non-magnetic}&\multirow{2}{*}{$D_{\text{Berry},xy}$}& $E_z$ induced $\kappa_{xx}$& $D_{\text{Berry},xy}\sim 10$ &$2$\,mW/(K\,m) \\ \cline{4-6}
  & &  &$\nabla_zT$ induced $\sigma_{xx}$ &  $ \Delta\mu\sim 10$\,meV & $2$\,S/m \\
 \hline\hline
  \multirow{4}{*}{$\text{SrSi}_2$ (Chiral Weyl SM)}&\multirow{4}{*}{Non-magnetic}& \multirow{4}{*}{Weyl chirality $\mathcal{S}_{tot}$}& Sum of the following: \newline $E_x$ induced $\kappa_{yz}$, \newline $E_y$ induced $\kappa_{zx}$,\newline $E_z$ induced $\kappa_{xy}$ &\multirow{4}{*}{ $T=100K$}  & \multirow{4}{*}{$0.24$\,mW/(K\,m)}\\
 \hline\hline
  \multirow{2}{*}{$\text{MnBi}_2\text{Te}_4$} & \multirow{2}{*}{Magnetic}&  \multirow{2}{*}{$D_{\text{Metric},yxx}$}& $E_x$ induced $\kappa_{xy}$ & $\sigma_{\text{INH},yxx}\sim 10$\,mA/V$^2$ & $0.25\,\mu$W/(K\,m) \\ \cline{4-6}
    & & & $\nabla_xT$ induced $\sigma_{xy}$  &$\Delta \mu\sim 0.1$\,eV & $30$\,$\mu$S/m \\
 \hline
\end{tabular}
    \caption{Candidate materials for experimental observation of nonlinear thermoelectric effects. In the estimation we have used relaxation time $\tau=10ps$, $\mathcal{E}=100V/m$, $\nabla T=1K/mm$ throughout. We have also used $T=10K$ throughout with the only exception being the chiral Weyl SM case with $T=100K$. In estimation we have used Sommerfeld expansions Eq.\eqref{eq:L_Mottformula},\eqref{eq:L_WiedemannFranzlaw},\eqref{eq:M_Mott_formula},\eqref{eq:M_Wiedemann-Franz_Law}, which enable us to infer the value of nonlinear thermoelectric responses from the existing nonlinear Hall measurements or DFT calculations. In estimating $\hat{L}^{112}$ and $\hat{M}^{112}$ based on Eq.\eqref{eq:L_Mottformula},\eqref{eq:M_Mott_formula}, we need assumptions on the energy derivative of BCD/QMD, which is characterized by a typical energy scale $\Delta \mu$ over which $D_{\text{Berry}}/D_{\text{Metric}}$ varies significantly. We estimate this energy scale based on existing DFT calculations on band structures.
    }
    \label{table:experiments}
\end{table*}

\section{Discussion}\label{sec:discussion}
In this work, we derived nonlinear thermoelectric responses induced by the Berry curvature dipole [Eqs.\eqref{eq:L112} and \eqref{eq:L212}] and the quantum metric dipole [Eqs.\eqref{eq:M112} and \eqref{eq:M212}], providing a theoretical framework for probing quantum geometry through nonlinear thermoelectric measurements. We used this framework to identify a quantized nonlinear thermoelectric effect that directly detects the Weyl node chirality (Eq.~\eqref{eq:thermal_chirality}).

Candidate materials for observing nonlinear thermoelectric effects include Weyl semimetals and topological insulators near topological phase transitions\cite{hasan2010colloquium,qi2011topological,armitage2018weyl}, where the quantum metric and Berry curvature diverge near Weyl nodes or band-touching points, leading to potentially large effects. As there are already plenty of DFT calculations and experiments on the BCD and QMD induced nonlinear Hall effects, we can utilize the Sommerfeld expansion to give a realistic estimate of the nonlinear thermoelectric signal in candidate materials. Promising candidates include transition metal monopinicide Weyl SMs such as NbP, NbAs, TaAs, TaP, and transition metal dichalcogenide Weyl SMs such as MoTe$_2$ and WTe$_2$, chiral Weyl SMs such as SrSi$_2$, CoSi, RhSi, and magnetic topological insulator MnBi$_2$Te$_4$\cite{gao2023quantum,wang2023quantum}.

As a closing comment let us focus on one especially promising candidate material, In-doped Pb$_{1-x}$Sn$_{x}$Te, and summarize other candidate materials in Table \ref{table:experiments}. For the purpose of making order-of-magnitude estimates, we use the following physical parameters throughout our estimation\cite{zhang2018berry}:
\begin{enumerate}
    \item The relaxation time $\tau$, which we take to be $10$\,ps.
    \item External electric field $\mathcal{E}$, which we take to be $\sim 10^2$\,V/m.
    \item Temperature gradient $\nabla T$, which we take as $\sim 1$\,K/mm.
\end{enumerate}

In-doped Pb$_{1-x}$Sn$_{x}$Te has an experimentally observed giant Berry curvature dipole $D_{\text{Berry},xy} \sim 10$ when the temperature is around $10$\,K \cite{zhang2021berry,zhang2022giant} and the chemical potential $\mu \approx 10$\,meV relative to the Weyl nodes. After the formation of ferroelectric order, the material has a polar axis along the $z$ direction. In order to observe the BCD-induced nonlinear thermoelectric effects, we can either measure the change of $\sigma_{xz}$ in the presence of $\nabla_x T$, or the change of $\kappa_{xz}$ in the presence of $\mathcal{E}_x$ (see Table.~\ref{tab:thermoelectric112} and Table.~\ref{tab:thermoelectric212} for other $D_{\text{Berry},xy}$ measurement setups). For our estimate, we assume that the material has $D_{\text{Berry},xy} = 10$ over an energy range $\sim 10$\,meV, so that the energy derivative $dD/d\mu\approx 1$\,meV$^{-1}$. Using the Sommerfeld expansion as given in Eq.~\eqref{eq:L_Mottformula}, we obtain a realistic estimate of the temperature-gradient-induced correction to electrical conductivity at $T = 10$\,K as follows:
\begin{equation}
    \hat{L}^{112}_{xzx}\nabla_x T\sim \frac{e^2}{\hbar^2}\frac{\pi^2}{3}k_B^2T\tau \left(2 \frac{dD_{\text{Berry},xy}}{d\mu} \right)(\nabla_xT)\sim 2\,\textrm{S/m},
\end{equation} 
which is a sizeable effect.

Similarly, using Eq.\eqref{eq:L_WiedemannFranzlaw}, we obtain a realistic estimate of the electric-field-induced correction to thermal conductivity at $T = 10$\,K as follows:

\begin{equation}
    \hat{L}^{212}_{xxz}\mathcal{E}_x\sim L_0T\frac{e^3\tau}{\hbar^2}D_{\text{Berry},xy}\mathcal{E}_x\sim 1\,\textrm{mW/(K\,m)},
\end{equation}
which is a measurable effect with available experimental techniques\cite{hirschberger2015thermal,vu2023magnon}.

Throughout this work we have limited our discussion to effects driven by the quantum geometry of the electron system. But phonons play a significant role in thermal conductivity, raising the natural question of whether they interfere with the effects we predict. One possible mechanism for phonon-induced electric-field dependence of $\kappa$ can be described as follows \cite{wooten2023electric}. In ferroelectric materials, the electric field affects the phonon dispersion, which in turn modifies the thermal conductivity along the field direction. Crucially, this mechanism is purely longitudinal and is fundamentally distinct from the transverse nonlinear responses derived in this work, which are rooted in Berry curvature effects of the electron system, with $\hat{L}^{212}_{aaa}$ and $\hat{M}^{212}_{aaa}$ identically vanishing. \textcolor{black}{For this reason a compelling demonstration of a quantum geometry-driven nonlinear thermoelectric effect may involve observing either a transverse response or a consistency between both a $E$-induced modification to $\kappa$ and a $(\nabla T)$-induced modification to $\sigma$ in the same material. Alternatively, one could examine the doping dependence of the effect across multiple samples, since doping strongly affects the BCD by changing the Fermi energy but has little influence on the phonon system. Preliminary measurements in In-doped Pb$_{1-x}$Sn$_x$Te observe a field-induced modification of the longitudinal thermal conductivity on the order of $1$ -- $10$ mW/(K\,m), as compared to an overall magnitude $\kappa \approx 2$ W/mK (which is phonon-dominated) \cite{heywood2026anomalous}. These measurements are consistent with our theoretical estimates above.}   Other more exotic phonon-related mechanisms might still emerge. We leave the exploration of these interesting possibilities for future study.

Finally, we note that nonlinear thermoelectric responses can be extended to other condensed matter systems, including superconductors \cite{watanabe2022nonreciprocal,tanaka2023nonlinear,hu2023supercurrent,matsyshyn2024superconducting} and magnonic systems \cite{matsumoto2011theoretical,mook2016tunable,su2017chiral,liu2019magnon}, providing new ways to experimentally explore quantum geometry in these materials.

\acknowledgments
We are grateful to J.~P.~Heremans, Y.-M.\ Lu, and N.\ Trivedi for useful discussions.
This work was supported by the Center for Emergent Materials, a National Science Foundation (NSF)-funded Materials Research Science and Engineering Center (MRSEC), under Grant No.\ DMR-2011876.

\

\textit{Data Availability}--The data that supports the findings of this article are openly available \cite{yang_dataset_2026}. 

\appendix
\section{Derivation of BCD-induced nonlinear thermoelectric coefficient}\label{appendix:derivation}
We provide a derivation of the BCD-induced nonlinear thermoelectric responses here. A similar derivation was provided in the high-energy context in Ref.~\cite{chen2016nonlinear} and in the condensed matter context in Ref.~\cite{nakai2019thermalelectric}.

The nonequilibrium distribution function $g_{\textbf{r},\textbf{k}}$ satisfies the following equation:
\begin{equation}
    \frac{\partial g_{\textbf{r},\textbf{k}}}{\partial t}+\dot{\textbf{r}}\cdot \nabla_{\textbf{r}}g_{\textbf{r},\textbf{k}}+\dot{\textbf{k}}\cdot\nabla_{\textbf{k}}g_{\textbf{r},\textbf{k}}=-\frac{g_{\textbf{r},\textbf{k}}-f_0}{\tau},
\end{equation}
where $\tau$ is the relaxation time. Assume the electric field is at frequency $\omega$: $\bm{E}(t)=Re[e^{i\omega t}\bm{E}_0]$, then we expand 

\begin{equation}
    g=f+g_{\nabla T}+g_{E\nabla T},
\end{equation}
where $g_{\nabla T}\propto \nabla T$, and $g_{E\nabla T}\propto \bm{E}\nabla T$. Keeping terms up to the same order, we can solve for $g_{\nabla T}$ and $g_{E\nabla T}$. First, for $g_{\nabla T}$, we have
\begin{equation}
    g_{\nabla T}=-\tau [\bm{v}\cdot \nabla T]\frac{\partial f}{\partial T}.
\end{equation}

Next, for $g_{E\nabla T}$, we have
\begin{equation}
    i\omega g_{E\nabla T}+\dot{\textbf{r}}\cdot\nabla_{\textbf{r}}f(\textbf{r})=-g_{E\nabla T}/\tau,
\end{equation}
where $f(\textbf{r})$ is in the local equilibrium with a position dependent temperature $T(\textbf{r})$. Inserting it into the equation of motion, we have: 
\begin{equation}
   g_{E\nabla T}=-\frac{\tau}{i\omega\tau+1} \frac{e}{\hbar}[(\bm{E}\times\bm{\Omega})\cdot\nabla T]\frac{\partial f}{\partial T}.
\end{equation}

The transport coefficient can be obtained from $\bm{j}_{e}=\int [dk]\int[dr](-e\dot{\textbf{r}})g$. Then $\hat{L}^{112}$ has two contributions from the second order nonlinear distribution function and the anomalous velocity. The result is:
%
\begin{align}
        \bm{j}_e = & \int[dk](-e\bm{v})\frac{1}{i\omega+1/\tau}\frac{e}{\hbar}[(\bm{E}\times \bm{\Omega})\cdot \nabla T]\frac{(\varepsilon-\mu)}{T}\partial_{\varepsilon}f+ \nonumber \\
        & (-e)\int[dk]\tau[\bm{v}\cdot \nabla T]\frac{(\varepsilon-\mu)}{T}(\partial_{\varepsilon}f)(\frac{e}{\hbar}\bm{E}\times\bm{\Omega}).
\end{align}

From this expression we can obtain the coefficient $\hat{L}_{abc}^{112}$:
\begin{align}
& \hat{L}_{abc}^{112}=\frac{e^2\tau}{\hbar}\int [dk]\frac{(\varepsilon-\mu)}{T}\left[\frac{\epsilon_{bcd}}{i\omega\tau+1}(\partial_af)\Omega_d-\epsilon_{abd}(\partial_cf)\Omega_d\right].
\end{align}

Similarly, from the definition of heat current $\bm{j}_Q=\bm{j}_E-\mu \bm{j}_e$, we have $\bm{j}_Q=\int [dk]\int[dr] [(\varepsilon-\mu)\dot{\textbf{r}}]g$ and the expression for $\hat{L}_{abc}^{212}$ is:
\begin{align}
& \hat{L}_{abc}^{212}=-\frac{e\tau}{\hbar}\int [dk]\frac{(\varepsilon-\mu)^2}{T}\left[\frac{\epsilon_{bcd}}{i\omega\tau+1}(\partial_af)\Omega_d-\epsilon_{abd}(\partial_cf)\Omega_d\right].
\end{align}

\section{Robustness of quantized thermoelectric effect}\label{appendix:robustness}
In this section, we provide more details on the robustness of the quantized nonlinear thermoelectric response against interactions and band curvature. 

Following the topological Fermi-liquid framework \cite{haldane2004berry,haldane2004ykis,wang2026interaction}, the inverse of the exact one-particle Green's function $G^{-1}(\mathbf{k}, \omega=0)$ is Hermitian at zero temperature ($T=0$). This allows us to define an effective single-quasiparticle Hamiltonian right at the Fermi surface:
\begin{equation}
\mathcal{H}^{qp}_{ij}(\mathbf{k}) = -\lim_{\omega \to 0} G^{-1}_{ij}(\mathbf{k}, \omega; T=0).
\end{equation}
From this interaction-renormalized Hamiltonian, we can self-consistently determine the Fermi surface topology, the quasiparticle Bloch states, and the renormalized Fermi velocities:
\begin{align}
&\mathcal{H}^{qp}(\textbf{k}_F) \ket{u^{qp}_{\textbf{k}_F}} = 0, \\
&\bm{v}_F(\textbf{k}_F) = \frac{1}{\hbar} \nabla_{\textbf{k}} \mathcal{H}^{qp}(\textbf{k})\Big|_{\textbf{k}=\textbf{k}_F}.
\end{align}

A key distinction between a strongly interacting Fermi liquid and a free electron gas is that well-defined quasiparticles only exist strictly on the Fermi surface. Consequently, the Berry connection computed from these quasiparticle Bloch states $\bm{\mathcal{A}}^{qp}(\textbf{k}) = -i \bra{u^{qp}_{\textbf{k}}} \nabla_{\textbf{k}} \ket{u^{qp}_{\textbf{k}}}$ and its corresponding Berry curvature $\bm{\Omega}^{qp}(\textbf{k})$ are only defined via derivatives parameterizing the 2D Fermi surface manifold. 

When calculating the nonlinear transport coefficients via the semiclassical Boltzmann equation, the response is governed by the geometric property of the Fermi surface. The corresponding Berry curvature dipole integral simplifies as follows:
\begin{align}
&\text{Tr}(D_{\text{Berry}}) =\hbar \int [dk] \delta(\varepsilon_F - \varepsilon(\textbf{k})) \bm{v}_F(\textbf{k}) \cdot \bm{\Omega}^{qp}(\textbf{k}) \nonumber \\
&= \oiint_{\text{FS}} \frac{d^2k_{\parallel}}{(2\pi)^3} \frac{\nabla_{\textbf{k}}\varepsilon \cdot \bm{\Omega}^{qp}}{|\nabla_{\textbf{k}}\varepsilon|}= \frac{1}{8\pi^3}\oiint_{\text{FS}} \bm{\Omega}^{qp} \cdot d\bm{S}=\frac{1}{4\pi^2} N.
\end{align}

Crucially, the density-of-states factor $1/|\nabla_{\textbf{k}}\varepsilon|$ exactly cancels the norm of the quasiparticle velocity, leaving only the unit vector normal to the Fermi surface. The resulting expression is therefore purely geometric and independent of microscopic band-structure details. Moreover, since the flux of the Berry curvature through a closed two-dimensional Fermi surface is a topological invariant associated with the homotopy class $\pi_2(S_2)$, the integral is strictly quantized and yields the Chern number of the Fermi surface.

At low temperatures, the nonlinear thermoelectric response receives contributions from states within a thermal window of width $k_B T$ near the Fermi surface. In a Landau Fermi liquid, however, the quasiparticle scattering rate vanishes as $\tau^{-1} \sim  T^2$, meaning that quasiparticles become asymptotically long-lived as $T \to 0$. Consequently, transport is governed entirely by quasiparticle states in the immediate vicinity of the Fermi surface. Provided that $k_BT\ll\varepsilon_F$, thermal broadening remains confined to a narrow shell around the Fermi surface, and the Fermi-liquid description remains valid. Therefore, so long as adiabatically turning on interactions does not induce a Lifshitz transition that alters the topology of the Fermi surface, the quantization of the nonlinear thermoelectric coefficient remains robust.

\vspace*{1cm}
\bibliography{thermal} 

\end{document}